\def\beq{\begin{equation}}
\def\eeq{\end{equation}}
\def\bea{\begin{eqnarray}}
\def\eea{\end{eqnarray}}
\def\bq{\begin{quote}}
\def\eq{\end{quote}}

\def\bseq{\begin{subequation}}  
\def\eseq{\end{subequation}}
\def\bsea{\begin{subeqnarray}}  
\def\esea{\end{subeqnarray}}

\def \gsim{\mathrel{\vcenter
     {\hbox{$>$}\nointerlineskip\hbox{$\sim$}}}}

\def\gappeq{\mathrel{\rlap {\raise.5ex\hbox{$>$}}
{\lower.5ex\hbox{$\sim$}}}}

\def\lappeq{\mathrel{\rlap{\raise.5ex\hbox{$<$}}
{\lower.5ex\hbox{$\sim$}}}}

\def\bbz{fa Z \kern-8.9pt Z}

\documentstyle [12pt]{article}

\begin{document}

\title{Model independent constraints on leptoquarks from rare processes}
\author{ Sacha Davidson \\ Center for Particle Astrophysics,
University of California\\ Berkeley, California, 94720 \\ \\ David
Bailey \\Department of Physics, University of Toronto\\ Toronto,
Ontario, M5S 1A7 \\ \\Bruce  A. Campbell \\Physics Department,
University of Alberta\\ Edmonton, Alberta, T6G 2J1}

\maketitle

\renewcommand{\arraystretch}{2}

\begin{abstract}
We present  model independent
 constraints on the masses and  couplings to fermions of   $B$
and $L$ conserving leptoquarks.  Such vector or scalar particles
could have masses below 100 GeV and be produced at HERA; we list the
generation dependent   bounds that can be calculated from rare
lepton and meson decays, meson-anti-meson mixing and various
electroweak tests.
\end{abstract}

\newpage
\section{Introduction}

\paragraph{}
A leptoquark is a scalar or vector particle that couples to a lepton
and a quark. It may or may not have well-defined  baryon and lepton number,
depending on the choices of coupling. Those with $B$ violating interactions
 would in general mediate proton decay, so their masses are expected to be
 very large ($\sim 10^{15}$ GeV). In this paper we only consider
leptoquarks with  $B$ and $L$ conserving renormalizable
couplings consistent with the symmetries of the Standard Model.

There are no interactions involving a quark, a lepton and a boson in
the Standard Model. There is a scalar Higgs doublet with electroweak
quantum numbers, and vector bosons that are either coloured or charged, but no
boson carrying colour and charge. This is a reflection of the fact that
 classically  the leptons and quarks appear to be independant unrelated
ingredients in the Standard Model. However, in each generation of
the quantum theory, they have equal and opposite contributions to the
hypercharge anomaly, which must vanish for the quantum theory to
make sense. It would therefore seem natural to have
interactions between the quarks and leptons in any extension of the
Standard Model, and, in consequence, bosons coupling to a lepton and
a quark.

At ordinary particle-anti-particle colliders, leptoquarks could be
 pair-produced, leading to a lower bound on their mass of roughly half the
 accelerator centre of mass energy. However, heavier ones can only be
exchanged in the $t$ channel, giving a small cross-section for
the interactions they mediate.  This problem is not present at
 a machine  colliding electrons and protons,
since a quark and an electron could form an $s$ channel leptoquark;
HERA, the $e$-$p$ collider at DESY, which has already
published some bounds on leptoquarks \cite{ZEUS}, should
be able to  see \cite{LQHERA,BRW} those with $m_{lq} < 314$ GeV (the
collider centre-of-mass energy) as peaks in the $x$ distributions of
inclusive processes, if such leptoquarks exist. They also expect to
see virtual effects due to  heavier leptoquarks \cite{BRW,ML2,virlq}.

 Present and
upcoming results that could give constraints on leptoquarks were
discussed  in \cite{Heu}. At low energies, leptoquarks could induce
two-lepton two-quark interactions, like those mediated by the
electroweak four-fermion vertices. This suggests that the
leptoquark yukawa coupling  squared ($\equiv \lambda^2$), divided by the
mass squared  ($\equiv m_{lq}^2$) is at least as small as $G_F$. However, for
certain combinations of generation indices on $\lambda$,
 there should be much stronger bounds because
 leptoquarks could mediate interactions
forbidden in the Standard Model by lepton family number conservation
and the absence of flavour-changing neutral currents (FCNC).

In this paper, we concentrate on bounds
from existing accelerators, meson-anti-meson  ($K \bar{K}$,  $D \bar{D}$,
$B \bar{B}$) mixing, rare lepton and meson  decays, and a few electroweak
tests. Most of these, have,  of
course,  been previously calculated in various models,
\cite{D+E,S1,S2,MSW,BKZ,BW,Leff,BRW,CEEGN,E6,D+H,ML2}
so we wish to list the bounds in as complete and
model independent a fashion as possible.
We neglect bounds from CP violation, because these constrain
the imaginary part of the coupling, not its magnitude.

In the remainder of this section, we introduce the
interactions we are allowing  the leptoquarks to have. There is then a
short review of some theoretical models in which leptoquarks appear,
 followed by sections on pre-HERA accelerator mass bounds,
constraints from meson decays, meson anti-meson mixing, muon physics and $\tau$
decays, and a  section of bounds that are none of the above.

We will assume that the couplings are real, so we
neglect any constraints arising from  leptoquark contributions to CP
violation. (This simplifying assumption was not made in
\cite{S1,CEEGN,D+H,HMP,G,BF}.)
 We will discuss each
calculation in the text, and usually list the bounds  (without
generation indices). Most of the numerical upper bounds  are
collected in  the tables at the end.

We neglect  QCD corrections to all the rates that we calculate,
except $ b \rightarrow s \gamma$. This may be a poor approximation,
 but we would not expect this to change our
bounds on $\lambda^2$ by more than a factor of 2.

 There are seven  renormalizable $B$ and $L$ conserving
quark-lepton-boson couplings consistent with the
SU(3)$\times$SU(2)$\times $U(1) symmetries of the Standard Model
for both scalar and vector leptoquarks, and each coupling carries
generation indices for the two fermions (these are suppressed in equations
(\ref{I-I}) and (\ref{I-II})). The scalar and vector interaction
lagrangians are therefore \cite{BRW}
\beq
\begin{array}{rcl}
{\cal L}_S & = & \left\{ \right. ( \lambda_{LS_o} \bar{q}^c_L i\tau_2 \ell_L
+ \lambda_{RS_o} \bar{u}^c_R e_R) S_o^{\dagger} + \lambda_{R \tilde{S_o}}
\bar{d}_R^c e_R \tilde{S}_o^{\dagger}  + (\lambda_{LS_{1/2}}
\bar{u}_R \ell_L +
  \\ & &  \lambda_{RS_{1/2}}\bar{q}_L  i\tau_2  e_R) S_{1/2}^{\dagger} +
 \lambda_{L \tilde{S}_{1/2}}
\bar{d}_R \ell_L \tilde{S}_{1/2}^{\dagger}
 + \lambda_{LS_{1}} \bar{q}^c_L i\tau_2 \vec{\tau}
\ell_L \cdot \vec{S}_1^{\dagger}  \left. \right\} + h.c. \label{I-I}
\end{array}
\eeq
and
\beq
\begin{array}{rcl}
{\cal L}_V &  = & \left\{ \right. ( \lambda_{LV_o} \bar{q}_L \gamma_{\mu}
\ell_L + \lambda_{RV_o} \bar{d}_R \gamma_{\mu} e_R) V^{\mu \dagger}_o
 + \lambda_{R\tilde{V_o}} \bar{u}_R
\gamma_{\mu} e_R \tilde{V}^{\mu \dagger}_o  + (\lambda_{LV_{1/2}}
\bar{d}_R^c \gamma_{\mu} \ell_L  \\   & &
+ \lambda_{RV_{1/2}}\bar{q}_L^c \gamma_{\mu} e_R)V_{1/2}^{\mu \dagger}
+  \lambda_{L \tilde{V}_{1/2}}
 \bar{u}_R^c \gamma_{\mu} \ell_L  \tilde{V}^{\mu \dagger}_{1/2} +
\lambda_{LV_1}
\bar{q}_L \gamma_{\mu} \vec{\tau} \ell_L \cdot \vec{V}^{\mu \dagger} _1
\left.  \right\} + h.c. \label{I-II}
\end{array}
\eeq
where the SU(2) singlet, doublet and triplet leptoquarks
respectively have subscipts 0, 1/2 and 1, and the
$L/R$ index on the coupling reflects the lepton chirality.
 We have written these
lagrangians in the so-called ``Aachen notation''. In the first column
of tables 3 and 4, we have written the interactions in this
notation, and also \{in curly brackets\}, in the notation of
reference \cite{BRW}. (In this alternative formulation,
fermion number two leptoquarks have  couplings $g$,
fermion number 0 leptoquarks have couplings $h$, and
SU(2) singlet, doublet and triplet leptoquarks
respectively have subscipts 1, 2 and 3.)
 The fermions in our  notation are chiral: here
and in the tables, $\bar{f}_R = (Rf)^{\dagger} \gamma^o$ and $f_L^c = (Lf)^c$.

 Generation
indices are superscripts and the lepton family index comes first:
$\lambda^{ij}$ couples a leptoquark to an $i$th generation lepton
and a $j$th generation quark. If leptoquarks
also carry generation indices,
cancellations between the different generations of leptoquark
are conceivable (vaguely similiar to the GIM mechanism).
In this case our bounds would apply  to the sum over
 leptoquark-induced amplitudes, but not  to
 the coupling constants and masses
of each generation of leptoquark.We ignore this possibility.
   In the sections on rare meson decays, we
 quote bounds on $e^2$ (see  (\ref{M1a}) and (\ref{M1b})), defined as
$\lambda^2
$ for vectors and $\lambda^2/2$ for scalars.

\section{Theory}

\paragraph{}
We present here a short review of some extensions of the Standard Model
that could contain leptoquarks. The first possibility is Grand Unified Theories
(GUTs) \cite{Ross}, where leptons and quarks usually appear in the
 same multiplet. In consequence
 it is quite normal to have gauge and Higgs
bosons coupling to a quark and  a lepton (with or without $B$ and $L$
violation). Secondly, there are extended technicolour theories\cite{TC},
 where quarks and leptons individually
 appear in  multiplets of the dynamically broken extended
technicolour group. The other particles in each multiplet are new fermions,
that would appear at low energies in fermion-anti-fermion
 bound states, some of
which are leptoquarks. And finally
in substructure models \cite{subrev}, the ``preons'' in a quark and lepton
could combine to form a scalar or vector leptoquark.

\subsection{Grand Unified Theories}

\paragraph{}
Leptoquarks first appeared in Pati and Salam's SU(4) model \cite{P+S}, where
lepton number was treated as a fourth colour: the four weak doublets of each
generation are arranged
 as a four of SU(4). SU(4) is then spontaneously
broken, so that the gluons remain massless and the leptoquark gauge bosons
become heavy. In this model, the leptoquarks induce flavour-changing neutral
currents and lepton family number
violation; they can also violate $B$, but their contribution to proton decay
is severely suppressed \cite{Lan}.

 Constraints on Pati-Salam type leptoquarks have been discussed in
\cite{P+S} and \cite{S1,S2}. One would expect these vector leptoquarks to have
 full-strength couplings to a lepton and a quark of the same generation,
giving, for instance, a large ($\sim \lambda^2/m_{lq}^2, \lambda
 \sim 1$) contribution
to $K \rightarrow e \mu ~~(s + \mu \rightarrow d + e)$. They could also
interact
 with quarks and leptons of different generations, but one would expect this to
be Cabbibo suppressed. So the best bounds on these leptoquarks should
come from interactions involving lepton-quark pairs  of the same generation.
 SU(5) \cite{G+G} GUT models  contain vector leptoquarks,
 but these acquire grand
unification scale masses when the gauge group breaks to $SU(3) \times
SU(2) \times U(1)$, so are of little interest to us. This is a generic
feature of GUTs: the quarks and leptons share multiplets, so there
 are necessarily leptoquark gauge bosons; however the lepton-quark
symmetry is broken at the GUT scale, so these gauge bosons have
large masses. (Note, however, that these are not neccessarily of order
$10^{15}$ GeV; the ``GUT-scale'' for Pati-Salam could be $\sim 10^5$ GeV.)
 SU(5) also contains scalar leptoquarks  in the Higgs sector.  The Standard
Model Higgs doublet that breaks SU(2) shares an SU(5) {\bf{5}} with an SU(2)
singlet leptoquark.  Given the
 fermion representation in SU(5), this leptoquark
necessarily has $B$-violating interactions because it couples
 to two quarks and also to a lepton and a quark.
 The scalar potential must therefore be arranged to give these leptoquarks
very large masses, in which case they are of no interest to us. Alternatively,
one can break SU(5) invariance by setting the leptoquark-quark-quark
couplings to zero, as was done in reference \cite{BW}, which allows all five
 components of the scalar {\bf 5} to be light. It is of course possible to have
scalar leptoquarks in SU(5) with $B$ and $L$ conserving interactions by putting
them in a representation other than the {\bf 5}. An example of this can be
found in \cite{M+Y},where it is shown that   the minimal SU(5) model plus
a light ($m_{lq} \sim 100$ GeV) SU(2) doublet leptoquark would be
compatible with present experimental results on proton decay
and $\sin^2 \theta_W$.  This $B$ and $L$
conserving leptoquark ($\equiv \tilde{S}_{1/2}$,
see table 3) occupies the same position in the SU(5) {\bf 10} as the quark
doublet, so its interactions with light Standard Model fermions are of
the form $\bar{d}_R l_L \tilde{S}_{1/2}^{\dagger}$.
 The other members of the {\bf 10} are assumed to be very heavy.

The field theory limit of the heterotic superstring is some
``GUT-like''  gauge field theory, which is broken at or below
the compactification scale  to the Standard Model and possibly some
extra $U(1)$s: $SU(3) \times SU(2) \times U(1)^{n+1}$. This low
energy limit may contain leptoquarks for the same reasons that GUTs do.
 For instance, certain Calabi-Yau compactifications have $G = E_6/H$
as the four-dimensional gauge group below the compactification scale, where $H$
is a discrete symmetry group that can be chosen such that $G$ is the
Standard Model  ($\times U(1)^n$). The low energy fields are then
 expected to sit in the {\bf 27} of $E_6$,

and for each generation consist  of quarks and
leptons, a right-handed neutrino, the two Higgs doublets of the Minimal
Supersymmetric
Standard Model, an extra SO(10) singlet, and a pair of SU(2) singlets,
which may have thr couplings of either diquarks or leptoquarks, but not
both simultaneously \cite{CEEGN}, as this would destabilize the proton.
The constraints on leptoquarks in such superstring-derived models
were calculated in \cite{CEEGN}, and are extensively reviewed in
 \cite{E6}.

\subsection{Technicolour}

\paragraph{}
If our knowledge of quantum field theory is extrapolated  to energies far
above those experimentally accessible today, one discovers that the masses
of fundamental scalars are generically the same size as the cutoff energy.
So to keep the mass of the Higgs at the electroweak scale, one must either
fine-tune the parameters of the theory to many decimal places, or introduce
 new physics that explains the comparatively small scalar mass. Both
supersymmetry and technicolour \cite{TC} do this.

In a technicolour theory, all scalars are bound states of fermion anti-fermion
pairs, like the mesons of QCD. One introduces new electroweak doublets and
singlets (technifermions) which are multiplets of
 a non-abelian gauge interaction called
technicolour. The new interaction
 becomes strong at the electroweak scale, and, by
analogy with QCD, one expects a techni-fermion condensate to form, breaking
some of the global symmetries  of the theory. Three of the goldstone bosons
arising from this global symmetry
breaking become the longitudinal components of the
electroweak gauge bosons. This neatly gives masses to the W and the Z, without
 a fundamental scalar Higgs. In fact, if one temporarily ignores the problem
 of generating fermion masses, one can simply introduce an electroweak doublet
of left-handed technifermions, and their right-handed singlet counterparts.
 This has an $SU(2)_L \times SU(2)_R$  global symmetry if one neglects the
(comparatively weak) standard model interactions. When the technifermion
condensate forms, the global symmetry is assumed to be
broken to $SU(2)_V$, and the three
goldstone bosons associated with the broken axial currents ($\sim$ pions in
QCD) are eaten by the electroweak gauge bosons.

To be a viable substitute for  the scalar Higgs, technicolour must also
produce fermion masses; this is done in extended technicolour (ETC) models.
 The idea is to introduce more technifermions, and gauge interactions
involving ordinary and technifermions that are broken at very high energies.
 The gauge bosons of these new interactions (which we refer to as  ETC gauge
bosons) presumably acquire mass by some dynamical symmetry breaking mechanism
of their own, and afterwards induce four-fermion interactions involving two
technifermions and two ordinary  fermions. After the formation of the
technifermion condensate, the four-fermion interaction becomes
a ``standard model'' mass term.

It is desirable to introduce a whole standard model generation of
 technifermions so that the ETC gauge bosons do not have to carry $SU(3)
 \times SU(2) \times U(1)$ quantum numbers. This means that there will be
``leptoquark technimesons'' made up of a techniquark and an anti-technilepton.
It is easiest to see why leptoquarks naturally arise in ETC models by looking
 at a simple example. An extra generation, plus right-handed neutrino, who
each transform as an {\bf N} under the technicolour SU(N), are added to
the  Standard Model, and one removes the Higgs sector. One also must add
other interactions and particles (including the ETC gauge bosons)
at higher energies. The new technifermions
are assumed to interact with each other via the techniforce, and with other
Standard Model particles via colour and electroweak interactions, and through
the effective four-fermion vertices induced by ETC gauge bosons. Technicolour
becomes strong around  1 TeV, a technifermion condensate is assumed to form,
and some of the global symmetries of the technicolour sector are dynamically
broken.

        In QCD,
if one neglects the masses of the $u$ and $d$ quarks, there is a global
$SU(2)_L \times SU(2)_R$ symmetry.
This is broken by the condensate to $SU(2)_V$,  and
the pions are the (pseudo-)goldstone bosons of the broken axial symmetry.
 Similar behaviour is expected in a technicolour theory; in the model
considered here, if one neglects electroweak, colour and four-fermion
interactions, there is a global $SU(8)_L \times SU(8)_R$ symmetry (assuming
complex representations under the technicolour group) which is expected to
break to  $SU(8)_V$. This produces 63 goldstone bosons. Three will be eaten
by the electroweak gauge bosons, and the others acquire masses due to the
neglected interactions. It is clear that among the 60 goldstone bosons, there
will be ``leptoquark mesons'' composed of a techniquark and an
anti-technilepton, bound together by the technicolour force. This scalar
leptoquark will interact with a quark and an anti-lepton carrying the same
Standard Model quantum numbers (it would be easy to construct a model where
 the leptoquark had fermion number 2).

The coupling of a technimeson leptoquark
to a lepton and a quark occurs when the technifermion constituents of the
leptoquark exchange an ETC gauge boson and turn into an ordinary quark and
lepton. The ``Yukawa'' coupling can only be predicted in a specific model,
 but it is expected to be of order $m_l/\Lambda$ or $m_q / \Lambda$, where
$\Lambda$ is the technicolour scale. Constraints on leptoquarks in
 technicolour models were computed in \cite{S1,S2,D+E}.

\subsection{Composite Models}

\paragraph{}
Leptoquarks frequently arise in composite, or substructure, models
\cite{subrev,AF,SUSYC,SS,other}.

In these scenarios, the leptons and quarks, and sometimes the
gauge bosons, are assumed to be bound states of more fundamental particles.
In principle, one could hope that this would explain the quark and lepton
spectrum, and reduce the number of fundamental fields (as quarks did for the
hadronic spectrum). However, at the moment, most models do not require fewer
 constituents (``preons'')
 than there are particles in the Standard
Model.

One of the fundamental questions in building theories where the Standard Model
 particles are composite, is to explain why their masses are much smaller than
the compositeness scale. A solution for the
fermion masses is to have them forbidden by a global chiral
symmetry in the strongly interacting sector of the theory. This symmetry
is broken by the Yukawa and SU(3)$_c \times$ U(1)  couplings, so one would
expect the masses generated by these symmetry-breaking effects to be small.
Models depending on preon chiral symmetries to
preserve the masslessness of composite fermion bound states must,
however, satisfy the 't Hooft anomaly matching conditions \cite{'tH}.
These require that chiral anomalies calculated in the effective
composite theory match those calculated in the preon
subtheory \cite{'tH,Frish,C+G}. A solution of the anomaly
matching conditions is to reinterpret the ``tumbling'' gauge theories
of dynamical symmetry breaking \cite{DRS1} as composite
models satisfying the 't Hooft conditions via the principle
of complementarity. Although solutions of the anomaly
matching conditions have been pursued by a variety of methods,
(for a review, see \cite{Peskin}), the construction of a
 fully realistic model remains an open problem. Another method
for keeping the composite fermions light is to make them
the supersymmetric partners of goldstone bosons in a theory
with a spontaneously broken approximate global symmetry.  Their
masses must then remain small even after the breaking of
supersymmetry and the global symmetry. It is even more difficult to
 explain how composite gauge bosons could have masses much below the
substructure scale; the Case-Gasiorowicz-Weinberg-Witten  theorem
forbids the presence of interacting massless spin-1 composite
 particles, suggesting that all composite vector
bosons should have masses of order the compositeness scale \cite{CGWW}.

If the quarks and leptons are composite particles, they should have form
factors.  The ``substructure contribution'' to $g-2$ of the muon
 is
expected to be of order $(m_{\mu}/\Lambda_c)^2$, where $\Lambda_c$ is the
compositeness scale.  This constrains $\Lambda_c \gsim 1$ TeV, which
 makes it
difficult to associate the weak scale with the substructure scale (as was done
in \cite{AF}) in realistic models.  There are a number of stronger, but more
model-dependant bounds on   $\Lambda_c$ \cite{PDB}.

Leptoquarks appear naturally in many substructure models\cite{subLQ}. This
 is not surprising: if  a constituent particle  of a Standard Model fermion
carries quark or lepton number, then a composite quark could turn into
a lepton in the presence of a composite lepton (that turned into a
quark) by exchanging the appropriate constituents.The bound
state consisting of the exchanged constituents would be a leptoquark. This
suggests that leptoquarks in composite models could naturally induce
interactions where  flavour or generation number was  conserved overall, but
 separately violated in the quark and lepton sector.

\section{Mass Bounds}

\paragraph{}
There have been numerous searches for leptoquarks at existing
accelerators. Possible searches at $e\bar{e}$ colliders were
discussed by Hewett and Rizzo \cite{H+R} who point out that the
measurement of the R-ratio from PEP and PETRA  constrains all scalar
leptoquarks to have $m_{lq} \geq$ 15 GeV. The leptoquarks are assumed
to be pair-produced in the decay of a virtual photon, so this bound
depends only on their electric charge which is known to be $\geq 1/3$
(see tables 1 and 2). The same measurement from AMY
\cite{AMY} gives
\beq
m_{lq} \geq 22.6 {\rm ~GeV ~~~~~~~ (for ~scalars).}   \label{I-III}
\eeq
We expect a similiar bound for vectors
\beq
m_{lq} \gappeq 20 {\rm ~GeV ~~~~~~~ (for ~vectors).}   \label{I-IV}
\eeq
These are the best model independant lower bounds on the mass (that we
are aware of). There have also been leptoquark searches  at LEP and
the $p \bar{p}$ colliders, but these look for leptoquark decay
products, so only apply if the leptoquark-lepton-quark Yukawa
coupling is ``sufficiently'' large (although this is
not a significant constraint; see below). The LEP lower bound \cite{L3} is
\beq
m_{lq} \geq 44 {\rm ~GeV}   \label{I-V}
\eeq
for scalars, and a similiar bound  should apply to the vectors. This
was calculated assuming that the $Z_o$ turned into a pair of
leptoquarks, which then decayed to jets and two leptons ($e \bar{e},
\mu \bar{\mu}, \tau \bar{\tau}, \nu \bar{\nu}$; the mass bound
actually depends slightly on the type of lepton
\cite{L3}---(\ref{I-V}) is an approximation). Since all the
leptoquarks are charged, they all couple to the $Z_o$, although this
is rather weak in the case of the $Q_{em} =  1/3$ scalar singlet (see
tables 1 and 2).

Delphi \cite{DDPF} has also looked for $Z$ decays into
an on-shell and an off-shell scalar leptoquark. They conclude
that
\beq
m_{lq} >  77~~~{\rm GeV}  ~~(\lambda > e)
\eeq
for leptoquarks decaying to a  quark and an electron or a muon.
Note that unlike the bounds (\ref{I-V}) and (\ref{I-aa}), this
only applies for large $(\lambda > e)$ leptoquark-lepton-quark
couplings.
In principle,
leptoquarks could contribute at one loop to the decays $Z_o \rightarrow
 \bar{\ell_1} \ell_2$, with $\ell_1 \neq \ell_2$, but the
 rates for such processes are so small that LEP would not expect to see
them even if the leptoquark yukawas were of order 1 \cite{Gen}.

The decay rate of a  leptoquark to a
lepton and a quark is
\beq
\Gamma = \frac{\lambda^2 m_{lq}}{16 \pi} ~~({\rm for ~scalars}) \label{D1}
\eeq
\beq
\Gamma = \frac{\lambda^2 m_{lq}}{24 \pi}  ~~({\rm for ~vectors}) \label{D2}
\eeq
so one would need the coupling constant to be very small ($\lappeq
10^{-8}$) for the leptoquark to decay outside a detector. Such a
small renormalizable coupling would be unexpected, so it should be
safe to assume $m_{lq} \geq 44$ GeV. (Note that equations (\ref{D1})
and (\ref{D2}) give the leptoquark decay rate to a specific flavour of
quark and lepton. The total decay rate would be the sum over all types
of kinematically accessible quarks and leptons.)

The  UA2 bound \cite{UA2} on scalar leptoquarks decaying to  $e +$
jets is $m_{lq} >$ 67 GeV, and CDF has a  bound
 of 113 GeV (95\%) \cite{Carl}.
 The strongest $\bar{p}p$ collider bound   is  a
  D0 result \cite{D0}  for leptoquark pairs  decaying to two
electrons plus jets:
\beq
m_{lq} > 126 {\rm ~~GeV~~~~~~(lq} \rightarrow e q ) \label{I-aa}
\eeq
where the leptoquark is assumed to decay entirely to
electron plus jets. The cross section for vector
leptoquark production is being calculated \cite{J+T}, so
bounds on vector leptoquarks from CDF and DO should be
 possible in the future.

It is not clear that leptoquarks that could be seen at HERA must satisfy
these bounds; although they must be produced in a quark-electron (or
positron) collision, they could have stronger couplings to muons or
taus, decay to $\mu$ or $\tau$ plus
 jets, and thereby avoid this constraint.
We will nonetheless use $m_{lq} = 100$ GeV in our numerical results.

The ZEUS and H1 Collaborations at HERA  have (mass-dependent)
limits on the couplings of scalar and vector
leptoquarks produced in an
$e p$ collision and decaying to $e + X$ or $\nu + X$ \cite{ZEUS}. The exact
bounds depend somewhat on the type of leptoquark, and
will of course improve with time. The present limits rule out roughly
from
\beq
       \lambda > .05 ~~~m_{lq} = 25 {\rm ~GeV}
\eeq
to
\beq
      \lambda > 1 ~~~m_{lq} = 225 {\rm ~GeV}
\eeq
for the SU(2) singlet and triplet leptoquarks (scalar and
vector), and from
\beq
       \lambda > .1 ~~~m_{lq} = 25 {\rm ~GeV}
\eeq
to
\beq
      \lambda > 1 ~~~m_{lq} = 150 {\rm ~GeV}
\eeq
for  the doublets. Note that these are just very rough approximations
to the present data. See \cite{ZEUS} for accurate plots.

\section{Meson Decays}

\paragraph{}
Since  leptoquarks have not been directly observed at any high energy
colliders, one must look for traces of them in other interactions
(meson decays, for instance). At energies well below their mass, the
leptoquarks can be ``integrated out'',   leaving effective
four-fermion vertices involving two leptons and two quarks, and
a photon-lepton-lepton (or quark-quark) magnetic moment
vertex.  We will come back to the latter in sections 6 and 7. The
four fermion vertices
contract the spinor indices of a lepton and a quark, but can be
Fierz-rearranged to contract the quark  indices together; in this
form one can estimate the quark matrix elements between initial and
final meson states. The leptoquark induced effective four-fermion
vertices are listed in tables 3 and 4.  The hermitian
conjugates of the listed operators are of course also possible. Most
of the leptoquarks couple to quarks of only one chirality, so the
Fierz-rearranged vertices must be of the form $V \pm A$, similiar to
the weak interactions. The only exceptions to this are $S_o, S_{1/2},
V_o^{\mu}$ and $V_{1/2}^{\mu}$, which have two interactions: a
``left-handed'' one with coupling constant $\lambda_L$,  involving
the leptoquark, a left-handed lepton and a quark, and a
``right-handed'' interaction coupling the leptoquark to a
right-handed lepton and a  quark. The Fierz-rearranged
effective  four-fermion vertex  arising when these two interactions
appear at opposite ends of the leptoquark propagator  contains
tensor, scalar and pseudoscalar pieces. We will ignore the tensor
operators, because their matrix elements are difficult to estimate,
and just list the scalar $\pm$ pseudoscalar parts.   Some
experimental results  will give much better constraints on the
coefficients of the $S \pm P$ matrix elements than the $V \pm A$
ones. In these cases, we will list bounds on the products $\lambda_L\lambda_R$.
In general, however, we will avoid the ``mixed coupling''
constraints because they do not apply if one of the interactions does
not exist.

In the Standard Model, many mesons are forbidden
from decaying via the strong or electromagnetic
interactions. Since such mesons decay at a rate consistent
with the weak interactions, one would
roughly  expect  the effective dimensionful coupling $\sim
\lambda^2/m_{lq}^2$  of the leptoquark four-fermion vertices to be as small
as that of the weak interactions:
\beq
\frac{\lambda^2}{m_{lq}^2}  \lappeq \frac{4
G_F}{\sqrt{2}} |V| \label{3.1}
\eeq
where $V$ is the appropriate CKM matrix element.

However, we can get much stronger constraints on the leptoquark
couplings that induce interactions not present in the Standard Model.
A variety of meson decay constraints were considered in references
\cite{BRW,ML2,S1,S2,MSW,BKZ,BW,CEEGN,D+H}.
The first ``non-Standard'' type of leptoquark-induced behaviour we
will consider is the lack of chiral suppression in meson decays due
to  pseudoscalar matrix elements. The V--A vertices of the weak
interactions give a matrix element squared for the leptonic decay of
pseudoscalar mesons proportional to $E_iE_n - \vec{k} \,^2$ (where
$E_i, E_n$ and $\vec{k}$ are the energies and momentum of the
outgoing leptons),
 because the helicity  ($\sim$ chirality for a
relativistic particle) of one of the outgoing leptons must be flipped
via a mass term to conserve angular momentum. The decay rate is
therefore strongly suppressed for
relativistic leptons. There is no such chiral
suppression in meson decays induced by leptoquarks with both left and
right-handed couplings (pseudoscalar matrix elements, see below).

 One
would also expect leptoquarks, if they exist, to induce lepton family
number violating decays; even if the leptoquark Yukawa interactions
are ``generation diagonal'' in the sense that they only couple first
generation quarks to first generation leptons, this would still
allow, for instance a kaon to decay to $\bar{e} \mu $. It would
similiarly be very natural for leptoquarks to induce flavour-changing
decays suppressed in the Standard Model.

   Leptoquarks can induce    effective four fermion vertices of the
form
\beq
\frac{e^{ij} e^{mn}}{m_{lq}^2} ( \bar{q}^j \gamma^{\mu} P ^{q}
q^n) ( \bar{\ell}^m \gamma_{\mu} P^{\ell} \ell^i)  \label{M1a}
\eeq
and/or
\beq
\frac{e^{ij} e^{mn}}{m_{lq}^2} ( \bar{q}^j  P^{q} q^n) (
\bar{\ell}^m  P^{\ell} \ell^i)  \label{M1b}
\eeq
(see tables 3 and  4)
where $P^q$ and $P^{\ell}$ are chiral projection operators =  L or R,
and $m_{lq}$ is the leptoquark mass. The coefficients of the
 effective operators in the third column of
tables 3 and 4 are generally twice as large for vectors as
for scalars. In equations (\ref{M1a})
and (\ref{M1b}) we  have introduced  a coupling $e^2$ =
$\lambda^2$ ($e^2$ = $\lambda^2/2$) for vectors (scalars);
the constraints  we  compute  on $e^2$  will  therefore  be
bounds on the square of the vector coupling, or half the square of the
 scalar coupling. As is clear from tables 3 and 4, $V \pm A$ quark
currents are induced by leptoquarks with couplings of the same chirality
at both ends of the propagator. The effective coupling of $V \pm A$
4-fermion vertices is therefore $e_L^2/m_{lq}^2$ or $e_R^2/m_{lq}^2$.
In all the following, we will quote bounds on $e_L^2$ with the
understanding that the constraint also applies to $e_R^2$. (Remember
that the coupling constant subscript applies to the
lepton chirality.) The
scalar and pseudoscalar vertices are induced by the
leptoquarks $S_o, S_{1/2},
V_o^{\mu}$ and $V_{1/2}^{\mu}$ with couplings of opposite
chiralities at either end of the propagator. The effective coupling
 of the $S \pm P$ vertices is therefore $e_L e_R/m_{lq}^2$.

  The matrix element squared for  the decay of a pseudoscalar meson $M $
($\bar{q}^j q^n$ bound state) to the leptons $\ell^i$ and
$\bar{\ell}^m$ is \beq
\begin{array}{lll}
|{\cal M}|^2  & = &  \frac{2}{m_{lq}^4} \{[( e_L^{ij} e_L^{mn})^2
+ ( e_R^{ij} e_R^{mn})^2] m_M^2
\tilde{A}^2(E_i E_m - \vec{k} \, ^2)  \\ &&+
[( e_L^{ij} e_R^{mn})^2  + ( e_L^{ij} e_R^{mn})^2]
\tilde{P}^2(E_i E_m + \vec{k} \, ^2 )   \\   & & +2
 (e_L^{ij} e_L^{mn} +e_R^{ij} e_R^{mn}) \tilde{P} \tilde{A} m_M (e_L^{ij}
e_R^{m
n}E_{i}
m_{m} -  e_R^{ij} e_L^{mn}E_{i} m_{m})\}
\end{array}
\label{M2}
\eeq
where \beq
\tilde{P} =\frac{1}{2} <0| \bar{q}^j \gamma^5 q^n |M> \sim \frac{f_M m_M}{2}
\frac{m_M}{m_{q^j} + m_{q^n}} \label{M3a}
\eeq
and
\beq
\tilde{A} \, p^{\mu} = \frac{1}{2}<0| \bar{q}^j \gamma^{\mu} \gamma^5
 q^n |M> =
\frac{f_M p^{\mu}}{2} \label{M3b}
\eeq
($p^{\mu}$ is the meson 4-momentum and $f_M$ the decay constant).
The approximation for $\tilde{P}$ is a current algebra result, so
 not  appropriate for the heavier  mesons; we will use it anyway to get
qualitative results.
Equation(\ref{M2}) gives a meson decay rate
\beq
\begin{array}{lll}
\Gamma(M \rightarrow \ell^i \bar{\ell}^m) &  = & \frac{k}{4 \pi m_M^2}
\frac{1}{m_{lq}^4}
 \{[( e_L^{ij} e_L^{mn})^2
+ ( e_R^{ij} e_R^{mn})^2] m_M^2
\tilde{A}^2(E_i E_m - \vec{k} \, ^2)  \\ &&+
[( e_L^{ij} e_R^{mn})^2  + ( e_L^{ij} e_R^{mn})^2]
\tilde{P}^2(E_i E_m + \vec{k} \, ^2 )   \\   & & +2
 (e_L^{ij} e_L^{mn} +e_R^{ij} e_R^{mn}) \tilde{P} \tilde{A} m_M (e_L^{ij}
e_R^{m
n}E_{i}
m_{m} -  e_R^{ij} e_L^{mn}E_{i} m_{m})\}
 \label{M4}
\end{array}
\eeq
where $k$ is the magnitude of the lepton 3-momentum in the centre-of-mass
frame, and the absence
of chiral suppression in the pseudoscalar matrix element is clear.

\subsection{Pion Decays}

\paragraph{}
In the Standard Model, charged pions decay principally to $\mu \nu$
because the decay to $e \nu$ is suppressed by angular momentum
conservation, as discussed in  the previous
section. The experimental ratio \cite{Rexpt}
\beq
R_{exp} \equiv \frac{\Gamma(\pi^+ \rightarrow \bar{e}
\nu)}{\Gamma(\pi^+ \rightarrow \bar{\mu} \nu)} = 1.231 \pm .006
\times 10^{-4} \label{M5}
\eeq
agrees with the Standard Model prediction \cite{Rth}
\beq
R_{th} = \frac{m_e^2}{m_{\mu}^2} \frac{(m_{\pi}^2 -
m_e^2)^2}{(m_{\pi}^2 - m_{\mu}^2)^2} (1 + \Delta) = 1.235 \pm .004 \times
10^{-4}  \label{M6}
\eeq
where $\Delta$ is a radiative correction.
 Using (\ref{M4}) to compute the  contribution to this
ratio from the interference of the Standard Model
axial vector amplitude and the presumably small
leptoquark-induced effective pseudoscalar operators, one can require
\beq
 \frac{\tilde{P}}{m_{\pi} \tilde{A}} \left( \right.
\frac{ e_L^{n1} e_R^{11}}{\sqrt{2} G_F m_{lq}^2} \frac{m_{\pi}}{m_{e}}
-\frac{ e_L^{n1} e_R^{21}}{\sqrt{2} G_F m_{lq}^2} \frac{m_{\pi}}{m_{\mu}}
\left. \right)
 < \frac{|R_{exp} - R_{th}|}{R_{th}} ~~. \label{M7}
\eeq
If we assume that there is no spurious cancellation between the
pseudoscalar contributions to $\pi \rightarrow \mu \nu$
and $\pi \rightarrow e \nu$ (first and second terms of equation (\ref{M7})),
then this gives
\beq
(e_L^{n1}e_R^{11})^2  < 5 \times 10^{-7} \left( \frac{m_{lq}}{{\rm 100 ~GeV}}
\right)^2 ~,~~~
 (e_L^{n1}e_R^{21})^2  <  10^{-4} \left( \frac{m_{lq}}{{\rm 100 ~GeV}}
\right)^2   \label{a}
\eeq
where we have used the current algebra result $\tilde{P} \simeq 7
f_{\pi} m_{\pi}$ \cite{S1}, and $n$ is a light neutrino index
($m_{\nu} \ll m_e$).

 We get bounds
on $(e_L^{11}/m_{lq})^2 $  from the interference between $W$ and
leptoquark exchange, which, if small, must satisfy
\beq
\frac{1}
{\sqrt{2}G_F m_{lq}^2} ( e_L^{11} e_L^{n1} - e_L^{21} e_L^{n1} ) <
\frac{|R_{th} - R_{expt}|}{R_{th}}
\eeq
where $n$ is again an arbitrary light ($m_{\nu} \ll m_e$) neutrino index.
Assuming that leptoquark couplings to muons and electrons are not
equal ($e_L^{11} \neq e_L^{21}$), we can conclude that:
 \beq
 e_L^{21}e_L^{n1} ,  e_L^{11}e_L^{n1} < 2 \times
10^{-3} \left( \frac{m_{lq}}{{\rm 100 ~GeV}} \right)^2 ~~. \label{b}
\eeq
If the leptoquark coupling is independant of the lepton generation, so that
$e^{11} = e^{21}$, then leptoquarks will  not affect
the $R$ ratio. However, $(e^{11})^2 [ = (e^{21})^2$ in this case], is still
constrained by quark-lepton universality (see section 8.1), which gives
a better bound than this one.

Equations (\ref{a}) and (\ref{b}) give bounds on the couplings of the
effective vertices of tables  3 and 4 that involve $u$,
$d$, $e$ (or $\mu$)
 and any flavour of light ($m_{\nu} \ll m_e$) neutrino.   We
list the exact constraints in the tables. Since the notation in
the tables is rather obscure, we will discuss  in detail
the first four rows of
the table listing bounds on vector leptoquark couplings (table 5).
The  four-fermion vertices have an effective
coupling constant $e^2/m_{lq}^2$, so the bounds on $e^2$ depend  on
$m_{lq}^2$. We take $m_{lq} = 100$ GeV to compute the numbers in the
tables, so the numerical bounds  scale as $(m_{lq}/100{\rm ~GeV})^2$.
The rows are labelled by experiments, and the columns by coupling
constants;  an entry in the table is the bound on a particular
coupling constant from the given experiment. These constraints  are
in general generation dependent, so we list the coupling constant
generation indices in parentheses.  For example, we have just found
that the upper bound on $R$ constrains axial-vector quark operator
couplings to be less than $2 \times 10^{-3} (m_{lq}/100{\rm~GeV})^2$. Only
leptoquarks inducing an interaction between $u,d,\nu$ and $e$ or $\mu$
could
contribute to $R$, so  we see from table 4 that we have bounds on
$\lambda_{LV_o}^2$ and $\lambda_{LV_1}^2$.
Since the neutrino in the decay $\pi ^+
\rightarrow \bar{e} \nu$  could be any species providing that it is
light, we  have the bounds
\beq
\lambda_{LV_o}^{11} \lambda_{LV_o}^{n1} <
2 \times 10^{-3} \left( \frac{m_{lq}}{100{\rm ~GeV}}
\right)^2
\eeq
\beq
\lambda_{LV_1}^{11} \lambda_{LV_1}^{n1} <
2 \times 10^{-3} \left( \frac{m_{lq}}{100{\rm ~GeV}}
\right)^2
\eeq
 where $n$ is the generation index of a light neutrino. This is the
first row of the  table. From (\ref{b}), the same bounds apply to
leptoquark couplings inducing the decay $\pi \rightarrow \mu \nu$,
which gives the second row of the table.  We also have bounds on pseudoscalar
operators from the $R$ ratio.
Again looking in table 4, we see  that there are such operators with
effective couplings $\lambda_{LV_o} \lambda_{RV_o}/m_{lq}^2$ and
$\lambda_{LV_{1/2}}
\lambda_{RV_{1/2}}/m_{lq}^2$. Checking
carefully in the second column to get the
right generation indices on the right coupling constants, we see that
the bound (\ref{a}) gives
 \beq
 \lambda_{LV_o}^{n1} \lambda_{RV_o}^{11} < 5 \times 10^{-7} \left(
\frac{m_{lq}}{100{\rm ~GeV}} \right)^2
\eeq
 \beq
 \lambda_{LV_{1/2}}^{n1}
\lambda_{RV_{1/2}}^{11} < 5 \times 10^{-7} \left(
\frac{m_{lq}}{100{\rm ~GeV}} \right)^2
\eeq
as listed in the third  row of the table. Note that bounds that cross
two columns apply to  the product of the couplings  in those columns,
with the first set of generation indices  applying to the first
column coupling constant and the second set to the second column.
The fourth row of the table  is the constraint from (\ref{a}) applied
to pseudoscalar operators involving muons.

{}From the upper bound BR($\pi^+ \rightarrow \bar{\mu} \nu_e) < 8
\times 10^{-3}$, one can constrain leptoquark induced effective
interactions of the form
\beq
\frac{e_L^2}{m_{lq}^2} ( \bar{u} \gamma^{\mu} P d)(
\bar{e}\gamma_{\mu} L \nu) ~,~~\frac{e_L e_R}{m_{lq}^2} ( \bar{u} P d)(
\bar{e} L \nu)
\eeq
 to satisfy
\beq
\frac{e_L^2}{m_{lq}^2}  < .25 G_F ~,~~ \frac{e_L e_R}{m_{lq}^2} < 5
\times 10^{-3} G_F \frac{m_{\mu}}{m_{\pi}}
\eeq
where we have again used $\tilde{P} = 7 f_{\pi} m_{\pi}$.

There is also an  upper bound on the lepton flavour
violating decay $\pi^o \rightarrow \mu^{\pm} e^{\mp}$: BR($\pi^o
\rightarrow \mu^{\pm} e^{\mp}$) $ < 1.6 \times 10^{-8}$,
and the branching ratio for $\pi^o \rightarrow e^{\pm} e^{\mp}$
has been measured \cite{pi0} to be $ \sim 7 \times 10^{-8}$.
Unfortunately, since the neutral pion decays electromagnetically
 $\pi^o \rightarrow \gamma \gamma$ with
a lifetime eight orders of magnitude smaller than that of the charged
pions, these do not translate into  very strong bounds on the
leptoquark couplings. We will get much better constraints, with the
same generation indices, from muon conversion on nuclei, and
atomic parity violation.

\subsection{Kaon Decays}

\paragraph{}
  Charged kaons decay  to $\mu \nu_{\mu}$ (BR = 63.5 \%) and a
variety of other final states, most involving pions. The decay to $e
\nu_{e}$ is suppressed by angular momentum conservation as in the
pion case, so if we require that the leptoquark contribution
to the branching ratio be less than the experimental error, we get the
results listed in the tables. We assume, as in the pion case, that there
is no cancellation between the leptoquark contributions to $K \rightarrow \mu
\nu$ and   $K \rightarrow e \nu$, and we use $\tilde{P} \sim f_K m_K/2$
to estimate the pseudoscalar contribution.

  If kaons decayed via a leptoquark, it would be just as natural to
have $\mu \nu_e$ in the final state as $\mu \nu_{\mu}$. The upper
bound on the ratio \cite{PDB}
  \beq
  \frac{BR(K \rightarrow \mu \nu_e)}{BR(K \rightarrow \mu \nu_{\mu})}
< 6.3 \times 10^{-3}
  \eeq
  implies
  \beq
  \frac{e_L^4}{8 m_{lq}^4 G_F^2 |V_{su}|^2} ~, ~ \frac{(e_L e_R)^2
\tilde{P}^2}{2 m_{lq}^4 G_F^2 |V_{su}|^2f_K^2 m_{\mu}^2} < 6.3 \times
10^{-3}
  \eeq
  where we  again use $\tilde{P} \sim m_K f_K/2$ to calculate the
numerical constraints listed in the tables.

  There are also bounds from the absence of the flavour-changing
decays $K \rightarrow \pi \ell_1 \bar{\ell}_2$, where $\ell_1 \bar{\ell}_2$
$ = e \bar{e}, \mu \bar{\mu}, \mu \bar{e}$ or
 $ e \bar{mu}$. The Standard Model does
not allow a strange quark to turn into a down quark at tree level
(FCNC), whereas leptoquarks could easily
induce an  $\bar{s} d \ell \bar{\ell}$ vertex. If we assume, using
isospin symmetry, that
 \beq
  <K^+ | \tilde{O} | \pi^o> = \frac{1}{\sqrt{2}}  <K^+ | \tilde{O} |
\pi^+>
  \eeq
  (where $\tilde{O}$ is some isospin 1/2 operator) and neglect all
lepton masses,  we can require
\beq
  \frac{e_L^4}{ m_{lq}^4} < 4 G_F^2 |V_{su}|^2  \frac{BR(K
\rightarrow \pi^+ \ell \bar{\ell})}{BR(K \rightarrow \pi^o \bar{\ell}
\nu)}
  \eeq
  giving the constraints  listed in   the tables. Neglecting the
masses should introduce an error of less than a factor of 2, because
$BR(K \rightarrow \pi^o \bar{e} \nu) \simeq 1.4 ~BR(K \rightarrow
\pi^o \bar{\mu} \nu)$.

   Finally we consider constraints  from leptonic decays of neutral
kaons. Most scalar and vector leptoquarks would allow the decays $K_L
\rightarrow \mu \bar{\mu}, e \bar{e}, \mu \bar{e}$, which are
suppressed in the Standard Model by the absence of flavour-changing
neutral currents (and lepton family number violation). There are both
axial vector and pseudoscalar quark matrix elements that can
contribute to these decays. We will neglect the pseudoscalar
contributions in the decays to $\mu \bar{e}$ and $\mu \bar{\mu}$
because they constrain the product $e_Le_R$, and the bounds on the
individual coupling constants ($e_L^2, e_R^2$) are only a factor of
$m_{\mu}/m_K$ weaker. Using equation (\ref{M4}) with $\tilde{A} =
f_K/2$ we find that
  \beq
  \Gamma(K_L \rightarrow e^{\pm} \mu^{\mp}) \simeq \frac{(m_K^2 -
m_{\mu}^2)^2}{64 \pi m_K^3} \frac{e_L^4}{m_{lq}^4} f_K^2 m_{\mu}^2 <
1.2 \times 10^{-27} {\rm ~GeV} \label{ref}
  \eeq
 which implies
 \beq
 e_L^2 < 6 \times 10^{-7} \left( \frac{m_{lq}}{100 {\rm ~GeV}}
\right)^2
 \eeq
 One can similiarly compute the decays $K_L \rightarrow \mu
\bar{\mu}$ and $ K_L \rightarrow e \bar{e}$ from equation (\ref{M4}).
This gives constraints as listed in the tables. We have included the
bounds from effective pseudoscalar operators in the decay to $e
\bar{e}$,  using $\tilde{P} \simeq m_K f_K/2$,  because they are
considerably stronger  (not suppressed by the small electron mass).
Note, however, that scalar leptoquarks do not induce pseudoscalar
operators of the form $(\bar{d^j} P d^m)(\bar{e^i} P e^m)$ (see
table 3), so these bounds only apply to $\lambda_{LV_o} \lambda_{RV_o}$ and
$\lambda_{LV_{1/2}} \lambda_{RV_{1/2}}$.

 \subsection{D decays}

 \paragraph{}
 Since the mass of the $D^+$ meson is much greater than that of the
muon, the decay $D^+ \rightarrow \mu^+ \nu$ is suppressed by angular
momentum conservation, like $\pi^+ \rightarrow e^+ \nu$. The present
upper bound on the  rate for $D^+ \rightarrow \mu^+ \nu$ is roughly a
factor of 2 larger than the Standard Model prediction, so is not the
best place to get bounds on leptoquarks  that induce axial vector
effective vertices. However, it does strongly constrain
effective pseudoscalar quark vertices
 involving $c, d, \mu$ and any neutrino:
 \beq
 \frac{(m_D^2 - m_{\mu}^2)^2}{16 \pi m_D} \tilde{P}^2
\frac{(e_L e_R)^2}{m_{lq}^4} < 4.5 \times 10^{-16} {\rm ~GeV}
 \eeq
 or, (assuming $\tilde{P} \simeq f_D m_D/2$, with \cite{Mart} $f_D
\sim .2 $ GeV)
 \beq
   e_L e_R < 6 \times 10^{-3} \left(\frac{m_{lq}}{100 {\rm ~GeV}}\right)^2 ~~.
 \eeq
A much weaker bound can be calculated for the effective coupling
of vertices involving $c$ and $s$  quarks, a muon and a neutrino,
from the upper bound $BR(D_S^+
\rightarrow \mu^+ \nu) < .03$. We do not list this in the tables.

 The CKM matrix element $V_{cd}$ is measured in $\Delta C = 1$ neutrino
interactions ($\nu + d \rightarrow \mu + c$) to be .204 $ \pm .017$ \cite{PDB}.
Requiring that leptoquarks contribute less than the total observed rate
(a conservative assumption) gives
\beq
 e_L^2 < \frac{4 G_F}{\sqrt{2}} |V_{cd}|  m_{lq}^2
 \eeq
for vertices involving $c, d, \mu $ and $\nu_{\mu}$. The upper bound
on the branching ratio for  $D^+ \rightarrow \mu^+ \nu$ gives slightly
weaker constraints, but applies to all  flavours of neutrino, so is
included in the tables.

 The CKM
matrix element $V_{cs}$ is determined to be $1.0 \pm .2$ from the
decay  $D^o \rightarrow K^- e^+ \nu$, so one can require
\beq
\frac{e_L^2}{m_{lq}^2} < \frac{4 G_F}{\sqrt{2}}
\eeq
where $e_L^2/m_{lq}^2$ is the effective coupling for an axial
vector  vertex involving
$c, s, e$, and any neutrino.

We can constrain the  effective coupling of the flavour-changing operator
($\bar{u}\gamma^{\mu}P c)(\bar{l}\gamma^{\mu}P l)$ ($l$ is an
electron or a muon) from the upper bound on the branching ratio
$BR(D^+ \rightarrow \pi^+ l \bar{l})$, if
we assume, by isospin symmetry, that
\beq
< D^+| \bar{u} \gamma^{\mu}P c|\pi^+> ~= ~< D^o| \bar{d} \gamma^{\mu}P c|\pi^->
  \eeq
(where $P = L,R$). The $D^o$ is observed to decay to  $\pi^- e \nu$
with a branching ratio of 3.9 $\times 10^{-3}$, so if we neglect all lepton
masses we can require
\beq
\frac{e_L^2}{m_{lq}^2} \frac{\sqrt{2}}{4 G_F |V_{cd}|} <
\sqrt{\frac{\tau_{D^o}}{\tau_{D^+}}\frac{BR(D^+ \rightarrow \pi^+ l \bar{l})}
{BR(D^o \rightarrow \pi^- \nu \bar{e})}}
\eeq
which gives the constraints listed in the tables.

  As in the case of kaons, one expects leptoquarks to induce
flavour-changing $D^o$ decays to lepton pairs ($ \bar{\mu} \mu,
\bar{e} e, \bar{\mu} e$). Most of the leptoquark interactions induce
effective axial vector quark and lepton currents, and since $m_D \gg
m_{\mu}, m_e$, the contribution of these currents to the decays $D
\rightarrow \bar{\mu} \mu, \bar{e} e, \bar{\mu} e$ is suppressed by
angular momentum conservation. One gets strong bounds on the
coupling constants of effective pseudoscalar   operators from the
 leptonic decays, but better bounds on the axial vector operators
from the previously considered $D^+ \rightarrow \pi^+ l \bar{l}$ decays.
  Using
(\ref{M4}) to calculate the rates for $D^o$ decay with the
approximation (\ref{M3a}),  the limit  BR($D^o \rightarrow \mu
\bar{e}$) $< 10^{-4}$ \cite{PDB} gives
  \beq
 e_L e_R < 4 \times 10^{-3} \left(
\frac{m_{lq}}{100 {\rm ~GeV}}\right)^2  {\rm ~~~ (for ~ } D^o
\rightarrow \mu \bar{e})
  \eeq
  Similiarly for BR($D^o \rightarrow \mu \bar{\mu}$) $< 1.1 \times
10^{-5}$ \cite{PDB} one gets
  \beq
 e_L e_R^2 < 10^{-3} \left(\frac{m_{lq}}{100 {\rm ~GeV}}
\right)^2 {\rm ~~(for ~ } D^o \rightarrow \mu \bar{\mu})
  \eeq
  and from
  BR($D^o \rightarrow e \bar{e}) < 1.3  \times 10^{-4}$ \cite{PDB}
  \beq
  e_L e_R^2 < 4 \times 10^{-3} \left(  \frac{m_{lq}}{100 {\rm ~GeV}}
\right)^2  {\rm (for ~ } D^o \rightarrow e \bar{e})
  \eeq

\subsection{B decays}

  One can get constraints on the leptoquark couplings to $b$ quarks
from the upper bounds on the flavour-changing decays $B \rightarrow
\ell \bar{\ell} X$    ($\ell \bar{\ell} = \mu \bar{\mu}, e \bar{e}$),
 $B \rightarrow
\ell_1 \bar{\ell}_2 K$  ($\ell_1 \bar{\ell}_2 = \mu \bar{\mu}, \mu \bar{e},
 e \bar{e}$),
and  $B^o \rightarrow \bar{\mu} \mu, \bar{e} e, \bar{\mu} e$.
If we approximate the hadronic matrix element in the  $B \rightarrow
\ell_1 \bar{\ell}_2 K$ decays via  heavy quark effective theory
\cite{MW}, we get some of our best constraints from these processes.
We nonetheless calculate and list the bounds from  $B \rightarrow
\ell \bar{\ell} X$ and  $B^o \rightarrow
\ell_1 \bar{\ell}_2 $ first because they should be more dependable.

The upper bounds on the leptonic branching ratios are
stronger than those for the inclusive  decays  $B \rightarrow
\ell \bar{\ell} X$, but the constraints on the couplings are weaker
because the theoretical rates
are suppressed by angular momentum conservation (small lepton masses).
 We therefore list,
for the leptonic decays, bounds on the coupling constants of
effective pseudoscalar vertices, and bounds on coupling constants
whose generation indices would allow $B^o \rightarrow \bar{\mu} e$,
because the latter are not constrained by the inclusive decay data.

  The decay $B^- \rightarrow e \bar{\nu} X$ ($B^- \rightarrow \mu
\bar{\nu} X$) is observed to have a branching ratio of 12 \% (11\%).
This is the rate expected in the Standard
Model with $|V_{cb}| = .05$,
 so we can require that the leptoquarks mediating this
interaction satisfy
\beq
\frac{e_L^2}{m_{lq}^2} \lappeq  \frac{4 G_F}{\sqrt{2}} |V_{cb}| \label{BB1}
  \eeq
The branching ratio for $B \rightarrow \tau \nu X$ has recently
been measured to be $4.2 +.72 -.68 \pm .46\%$ by Aleph \cite{HEPdal}.
One would expect the rate to be suppressed by about a factor of 2.5 with
respect
to $B \rightarrow \mu \nu X$ by kinematics, so leptoquarks
coupling a tau, a neutrino, a $b$ and a $c$ or $u$ quark
must also satisfy (\ref{BB1}).
  If we neglect differences in the kinematics between $B^+
\rightarrow \ell \bar{\ell} X$ and $B^+ \rightarrow \nu \bar{\ell} X$,
where $\ell = e, \mu$,
we can also constrain
  \beq
  \frac{e_L^4}{m_{lq}^4} <  8 G_F^2 |V_{cb}|^2 \frac{BR(B^+
\rightarrow \ell \bar{\ell} X)}{BR(B^+ \rightarrow \nu \bar{\ell} X)}
\label{BB2}
  \eeq
 (\ref{BB1}) and (\ref{BB2}) give  bounds on four
fermion vertices involving $e \nu$  $b$   $c$ (or $e \nu$  $b$ $u$,
 but we will get better bounds on this from (\ref{BB3}),
  $\ell \bar{\ell}bs$ and $\ell \bar{\ell}bd$.

Taking $V_{cb} = .06$, we obtain the constraints listed in the
tables. The ratio of CKM matrix elements $V_{ub}/V_{cb}$ is measured
to be \cite{Vub} .10 $\pm .03$ by fitting the lepton energy spectrum
in semi-leptonic $B$ decays. We can therefore constrain leptoquarks coupling a
$b, u, e$ and a $\nu$ to satisfy
\beq
\frac{e_L^2}{m_{lq}^2} \lappeq  \frac{4 G_F}{\sqrt{2}} |V_{ub}| ~~.
\label{BB3}
\eeq

We can get constraints  on four fermion interactions involving  $b$,
$\tau$, a down-type quark ($d$ or $s$) and another lepton ($\ell = e$, $\mu$ or
$\tau$) from the decays  $B \rightarrow
\ell \bar{\ell} X$ because the $\tau$ decays very rapidly. Suppose first that
$\ell = e$, and the leptoquark mediates the decay  $B \rightarrow
\tau \bar{e} X$. We are interested in the
$\tau$  decaying to $\nu$ + hadrons, in which case
this would contribute to  $B \rightarrow
\nu \bar{e} X$, or the tau decaying to $e \nu \bar{\nu}$, giving  $
e \bar{e} \nu \bar{\nu} X$ as the final state. The experimental
limit on the branching ratio for $B \rightarrow
e \bar{e} X$ is smaller than the measured rate for
$B \rightarrow  \nu \bar{e} X$, so we would get better constraints
if we could assume  $B \rightarrow
e \bar{e} \nu \bar{\nu} X \subset B \rightarrow  e \bar{e} X$. However,
the CLEO Collaboration \cite{CLEOt} required that the
$e\bar{e}$ pair in its upper bound on $B \rightarrow  e \bar{e} X$
come out back to back, and since the tau is not extremely relativistic,
it is not clear that the electron from its decay would be at 180$^o$ from
the $\bar{e}$.We will therefore calcalate bounds from the rate
for $B \rightarrow \nu e X$.
 We  assume that  the phase space suppression due
to the tau mass
 contributes about a factor of 1/2.5, so
\beq
\Gamma(B \rightarrow \tau e X) = \frac{1}{2.5}
\frac{e^4}{m_{lq}^4} \frac{\Gamma(B
\rightarrow e \nu X)}{8 G_F^2 |V_{cb}|^2}~~.
\eeq
 We  can therefore require that  couplings involving
$b, \tau, e$ and an $s$ or $d$ satisfy
\beq
\frac{e^2}{m_{lq}^2} < 2 \sqrt{5} G_F \frac{V_{cb}}{[BR(\tau
\rightarrow \nu X')]^{1/2}}~~.  \label{tau}
\eeq
If the $b$ decayed to $\tau, \mu$ and a $s$ or a $d$ quark, and the
$\tau$ decayed to $\mu \nu \bar{\nu}$, this would probably have
been seen by UA1 in their search \cite{UA1} for $B \rightarrow \mu
\bar{\mu} X$. We can therefore require
\beq
\frac{e^2}{m_{lq}^2} < \frac{4 G_FV_{cb}}{\sqrt{2}}
 \sqrt{\frac{BR(B \rightarrow \mu \mu X)}
{BR(B \rightarrow \mu \nu X)BR(\tau \rightarrow
\mu \nu \bar{\nu})}}
\eeq
for couplings involving $b, \tau, \mu$ and a $d$ or an $s$ quark.

Finally we consider the decays $B \rightarrow \tau \tau X$. We assume that
both taus decay to muons, and we include an
approximation to the the phase space suppression
due to the masses of the final state particles {\footnote{We
thank Andrzej Czarnecki for creating a simple analytic
approximation for us.}}. The effective coupling
of four-fermion vertices involving a $b$, two $\tau$s, and a $d$ or an $s$
quark must therefore satisfy
\beq
\frac{e^2}{m_{lq}^2} < \frac{4 G_F}{\sqrt{2}} \frac{V_{cb}}{BR(\tau \rightarrow
\mu \nu \bar{\nu})} \sqrt{\frac{BR(B \rightarrow \mu \mu X)}
{BR(B \rightarrow \mu \nu X)~PS}}
\eeq
where $PS$ is the phase space suppression factor:
\beq
PS = \left[ 1 - \left(\frac{2 m_{\tau}}{m_b} \right)^{2} \right]^{2.48}
{}~~.
\eeq

  The rate for the decay $B^o \rightarrow \bar{e} \mu$ via  an
effective axial vector quark operator will be (\ref{ref}) with the
obvious modifications:
  \beq
  \Gamma(B^o \rightarrow e^{\pm} \mu^{\mp}) = \frac{(m_B^2 -
m_{\mu}^2)^2}{64 \pi m_B^3} \frac{e_L^4}{m_{lq}^4} f_B^2 m_{\mu}^2 <
3 \times 10^{-18} {\rm ~GeV}
  \eeq
where we have used the recent  CLEO bound $BR(B^o \rightarrow
\bar{e} \mu) < 6 \times 10^{-6}$ to get the numerical constraint,
and we assume  $f_B \sim .2 $ GeV to get the constraints in the tables.
 Using  $\tilde{P}  \sim f_B m_B/2$ in (\ref{M4}), we can
estimate the decay via a pseudoscalar vertex  to be
  \beq
  \Gamma_P \sim .03 \frac{(e_L e_R)^2}{m_{lq}^4} {\rm ~GeV}^5 \lappeq 2
\times 10^{-17} {\rm ~GeV}
  \eeq
  which gives the bounds on $\lambda_{L} \lambda_{R}$ listed in the
tables. These
constraints, however, are only rough estimates.

 The most stringent experimental bounds on flavour-changing
neutral currents involving $b$ quarks come from the exclusive decays
$B^+ \rightarrow X^+ \ell_1^+ \ell_2^- $, where $\ell_1^+ \ell_2^- $
can be $e \bar{e}$, $\mu \bar{e}$ or $\mu \bar{\mu}$, and $X$
is a $K$ or a $\pi$. However, to calculate a
rate for this decay, we need to evaluate the matrix element
$<X^+|\bar{s} \gamma^{\mu} b|B^+>$. We
do this by relating  it at zero recoil to  $<X^+|\bar{s} \gamma^{\mu} c|D^o>$
via the heavy quark formulism \cite{MW}, and then making
very simple approximations for the momentum transfer dependence
of the form factors.

The $B$ to $X$ matrix element can be written
\beq
<X^+|\bar{s} \gamma^{\mu} b|B^+> = f_+^{B\rightarrow X} ( p_B^{\mu} +
p_X^{\mu} )  + f_-^{B\rightarrow X} ( p_B^{\mu} -
p_X^{\mu} )
\eeq
where $f_{\pm}$ are functions of the momentum transfer $t = ( p_B -
p_X)^2$. A similiar expression can be written for
the $D^o \rightarrow X^+$ matrix element, and the $B$ and $D$ form
factors can then be related by the heavy quark symmetry \cite{IW}.
Neglecting corrections of order $m_c/m_b$ and $\Lambda/m_b$, this gives
\beq
f_+^{B \rightarrow X} (t_{max}) = - f_-^{B \rightarrow X} (t_{max}) =
\sqrt{ \frac{m_b}{m_c}} \left[ \frac{\alpha(m_b)}{\alpha (m_c)} \right]^{-6/25}
f_+^{D \rightarrow X} (t_{max}) \label{hq1}
\eeq
where $t_{max}$ is the maximum momentum transfer between the
mesons ($(m_B - m_X)^2$ for the $B$ decays, $(m_D - m_X)^2$ for the
$D$s).

The decay rate for
$B \rightarrow X \ell^+ \ell^-$ is proportional to
$f_+^2$ \cite{ISGW}, so if we were to ignore the $t$
dependance of the form factors, and the mass of all
the final state particles in
both the $B$ and $D$  decays, we could require that
\beq
\frac{e^4}{m_{lq}^4} < 8 G_F^2 |V|^2 \frac{m_D^5}{m_B^5}
\frac{m_c}{m_b} \left[ \frac{\alpha(m_b)}{\alpha (m_c)} \right]^{12/25}
\frac{\Gamma(B^+ \rightarrow X \ell^+ \ell^-)}
{\Gamma(D^o \rightarrow X \ell \nu)}
\label{hq2}
\eeq
where $|V|$ is the appropriate CKM matrix element, and
$\Gamma(B^+ \rightarrow X \ell^+ \ell^-)$ is the experimental upper
bound on this rate.

These approximations are rather extreme; the form factors
can vary by up to an order of magnitude as $t$ varies from $t_{min}$ to
$t_{max}$, and $m_K/ m_D$ is far from negligable. To model the
momentum dependence of the form factors, we assume that
they  vary with $t$ as
\beq
f_+(t) = \frac{f_+(0)}{1 - t/m^2}
\eeq
(where we take the mass to be that of the decaying pseudoscalar
meson for convenience; the vector meson mass might be more accurate, but
for $b$ and $c$ mesons, these are very similiar), then
(\ref{hq1}) becomes (if we neglect $m_X/m_B, m_X/m_D$)
\beq
f_+^{B \rightarrow X} (0) = \frac{m_D}{m_B}
\sqrt{\frac{m_b}{m_c}} \left[ \frac{\alpha(m_b)}{\alpha (m_c)} \right]^{-6/25}
f_+^{D \rightarrow X} (0)
\eeq
and the bound (\ref{hq2}) is
\beq
\frac{e^4}{m_{lq}^4} < 8 G_F^2 |V|^2 \frac{m_D^3}{m_B^3}
\frac{m_c}{m_b} \left[ \frac{\alpha(m_b)}{\alpha (m_c)} \right]^{12/25}
\frac{\int d\Pi_3 (1 - t/m_D^2)^{-2} {\cal F}_D }
{\int d\Pi_3 (1 - t/m_B^2)^{-2} {\cal F}_B }
\frac{\Gamma(B^+ \rightarrow X \ell^+ \ell^-)}
{\Gamma(D^o \rightarrow X \ell \nu)}
\label{hq3}
\eeq
where the integrals are over final state phase space, and ${\cal F}$
is the appropriate integrand (which is independent of the decaying meson mass
in the approximation that
 final state masses are neglected).The ratio of integrals
will depend on the masses of the final state mesons: if
 the integrals are dominated
by $t \sim t_{max}$, then their ratio  could be
of order $m_D^2/m_B^2$, giving the constraint (\ref{hq2}). If instead,
the small momentum transfer region is dominant, the ratio should be $\sim 1.$
We can skirt these difficulties to some extent by comparing
$B^+ \rightarrow K^+ \ell^+ \ell^-$ to $D^o \rightarrow \pi^+ e \nu$
because $m_{\pi}/m_D \approx m_K/m_B$. If we assume
that $f_+^{D \rightarrow K}(0) =f_+^{D \rightarrow \pi}(0)$
\cite{eq}, then we get (\ref{hq2}) but with the
branching ratio for  $D^o \rightarrow \pi^+ e \nu$ and the
upper bound on   $B^+ \rightarrow K^+ \ell^+ \ell^-$. For the decays
$B^+ \rightarrow \pi^+ \ell^+ \ell^-$, we will use (\ref{hq3}),
assuming that the ratio of integrals is 1, because this
gives the weakest bounds. This series of approximations
gives the constraints listed in the tables, but we
emphasize that they are very sloppy.

CLEO has recently observed  the decay $B^* \rightarrow K^* \gamma$
with a branching ratio of $(4.5 \pm 1.5 \pm .9) \times
10^{-5}$ \cite{BKg}, which is consistent
with the rate expected from the Standard Model. Leptoquarks could also
contribute via the diagrams of figures 3 and 4, with $b$ and $s$ quarks
on the external legs, and a charged lepton $\ell$ in the loop. We can
therefore calculate bounds on  couplings $\lambda^{\ell s} \lambda^{\ell b}$
by requiring that the leptoquark contribution to $b \rightarrow s \gamma$
be ``sufficiently small''. The leptoquark contribution to
$b \rightarrow s \gamma$ in a string inspired supersymmetric model was
calculated in \cite{D}, but is not applicable to the simple
leptoquark case because of the presence of the supersymmetric partners.

In section 6.2, we calculate bounds on leptoquarks from the decay
$\mu \rightarrow e \gamma$. By making the initial fermion a $b$, the loop
fermion a lepton, and dividing by three (no colour sum)
in equations (\ref{m6b}),(\ref{m6}),(\ref{ope}) and (\ref{ngv2}),
  we can
calculate $\Gamma(b \rightarrow s \gamma)$ at a
 scale $\mu \sim m_{W} \sim m_{lq}$ from equation (\ref{m5}).
To get a rough idea of the magnitude of the bound on
leptoquarks from $b \rightarrow s \gamma$, we can simply
add the Standard Model and leptoquark amplitudes,
run the sum down to $m_{b}$ \cite{QCD}, and require that the total rate
fit into CLEO's allowed window. Since the renormalisation group
running mixes $b \rightarrow s \gamma$  with $b \rightarrow s G$, we need
the leptoquark contribution to $b \rightarrow s G$. Instead,
we will assume that it is smaller than the Standard Model
amplitude, and neglect it. With these approximations, the
 $b \rightarrow s \gamma$ amplitude at $m_b$ can be written
\beq
{\cal A} (b \rightarrow s \gamma) = a [ {\cal A}_{SM}
(b \rightarrow s \gamma) + {\cal A}_{LQ} (b \rightarrow s \gamma) + b]
\eeq
where $a$ and $b$ parametrize the QCD running. The relative sign
between the Standard Model and leptoquark amplitudes is
unknown, as is the exact value of the Standard Model amplitude (it depends
 on $m_t$), and the relationship between the measured rate for
$B \rightarrow K^* \gamma$ and the calculated one for $b \rightarrow
s \gamma$. So to get a rough idea of the magnitude of
the $b \rightarrow s \gamma$ bound on leptoquarks, we will
 require
\beq
 |{\cal A}_{LQ} (b \rightarrow s \gamma)| \lappeq
 |{\cal A}_{SM} (b \rightarrow s \gamma)|
 \eeq
where we take $m_t = 2 m_W$ in the Standard Model amplitude.

Following reference \cite{Drew}, we get for scalar leptoquarks:
\beq
\lambda_L^{\ell b} \lambda_L^{\ell s},  \lambda_R^{\ell b} \lambda_R^{\ell s}
< \frac{3 \times 10^{-2}}{Q_{\ell} + Q_{lq}/2}
\left( \frac{m_{lq}}{100 {\rm GeV}} \right)^2 ~~~({\rm scalars}) \label{ff1}
\eeq
where $\ell$ is a charged lepton, and $Q_{\ell}$ and $Q_{lq}$  are defined
as the quark and leptoquark electric charges coming out of the incident $b$
vertex.

There is a similiar bound for non-gauge vectors:
\beq
\lambda_L^{\ell b} \lambda_L^{\ell s},  \lambda_R^{\ell b} \lambda_R^{\ell s}
< \frac{ 2 \times 10^{-2}}{2Q_{\ell} + \frac{5}{2}Q_{lq}}
\left( \frac{m_{lq}}{100 {\rm GeV}} \right)^2 ~~~({\rm non-gauge\ vectors})
\label{ff2}
\eeq
where $2Q_{\ell} + \frac{5}{2}Q_{lq}$ = 1/3 (2/3) for $\lambda_{LV_o}$,
$\lambda_{RV_o}$ and $\lambda_{LV_1}$ ($\lambda_{LV_{1/2}}$,
$\lambda_{RV_{1/2}}$). We do not list the non-gauge  bounds in the tables. The
bounds on gauge vectors are to weak to be interesting.
Equations (\ref{ff1}) and (\ref{ff2}) are really just estimates of the
 magnitude of the bound that could be derived; however, since these
couplings are better constrained by other $B$ decays (see table 15),
this is not a serious problem.

\section{Meson-anti-meson mixing:$K^o$-$\bar{K}^o$, $D^o$-$\bar{D}^o$
and $B^o$-$\bar{B}^o$}

Weak interaction box diagrams (see figure 1) neatly account for the
small mass term that mixes the $K^o$ and $\bar{K}^o$ mesons.
 Similiar box diagrams, with leptoquarks instead of $W$ bosons and
the internal quarks replaced by leptons would also contribute to the
$K^o$-$\bar{K}^o$  mass difference, and one can therefore derive
bounds on the leptoquark couplings from these amplitudes.
This has been done in \cite{S1,BKZ,CEEGN,D+H} and more recently in \cite{ML}.

  The Standard Model boxes are GIM suppressed ( of order $G_F \alpha
(m_q/m_{W})^2$ rather than $G_F \alpha$) because the $qqW$ coupling
constant matrix in generation space is unitary. This may also be the
case for vector leptoquarks, if they are  the vector bosons of some
spontaneously broken gauge symmetry (this is the only renormalisable
way of making spin-1 particles; however, composite and technicolour
models do not neccessarily reduce to renormalizable low-energy
effective theories.).  The gauge boson leptoquark contribution  to
meson-anti-meson mixing would be of the same form as the
Standard Model one, giving upper bounds on the vector leptoquark
couplings of order (\ref{3.1}). If the vector leptoquarks are not
gauge bosons, GIM-type suppression is unlikely, and the box diagrams
for vectors would be of order $\lambda^4/m_{lq}^2$. This would give
stronger constraints and will be briefly discussed at the end of the
next  section. We assume in the tables however, that the vector
leptoquarks are gauge bosons, and list the weaker bounds.

   Scalar leptoquark box diagrams will of course be of order
$\lambda^4/m_{lq}^2$ and we should get  interesting constraints from these.

\subsection{Vector leptoquarks}

The Standard Model $W$ boxes contributing to the $K^o$-$\bar{K}^o$
mass difference give \cite{C+L}
\begin{eqnarray}
\frac{\Delta m_K}{2} & = & \frac{G_F}{\sqrt{2}} \frac{\alpha}{4 \pi
\sin^2 \theta_W} \left[  \right. \sum_{q = c,t} (V_{qs} V_{qd}^*)^2
\frac{m_q^2}{m_W^2} \\ \nonumber  & &  + V_{cs} V_{cd}^* V_{ts}
V_{td}^* \frac{2 m_c^2 m_t^2}{m_W^2 ( m_t^2 - m_c^2)} \ln \left(
\frac{m_t^2}{m_c^2} \right)  \left.\right] <K|(\bar{d} \gamma^{\mu} L
s)(\bar{d} \gamma_{\mu} L s)|\bar{K}> \label{K0}
\end{eqnarray}
where $V$ is the CKM matrix, and we have used the identity
$(\bar{q} \gamma^{\alpha} \gamma^{\mu} \gamma^{\beta} L
q')(\bar{q} \gamma_{\beta} \gamma_{\mu} \gamma_{\alpha} L q')$
 $ =$ $4(\bar{q} \gamma^{\mu} L
q')(\bar{q} \gamma_{\mu} L q')$. If one neglects the top quark
contributions because $V_{td} \ll 1$, and approximates the matrix
element  (vacuum saturation) as
\beq
<K|(\bar{d} \gamma^{\mu} L s)(\bar{d} \gamma_{\mu} L s)|\bar{K}>
\simeq \frac{1}{3} f_K^2 m_K        \label{K1}
\eeq
 this agrees quite well with the measured mass difference $\Delta m_K
= 3.5 \times 10^{-15}$ GeV. Gauge   leptoquarks will give the same
contribution to $\Delta m_K$, with $g_W/\sqrt{2}$ replaced by $ \lambda_{lq}$
and
lepton masses instead of quarks. We therefore require
\beq
\frac{1}{32 \pi^2 m_{lq}^2} \left[ \right. |\lambda^{ld} \lambda^{ls}|^2
\frac{m_l^2}{m_{lq}^2}  + \lambda^{ld} \lambda^{ls} \lambda^{l'd} \lambda^{l's}
\frac{2m_l^2m_{l'}^2}{m_{lq}^2(m_l^2 - m_{l'}^2)} \ln
\left( \frac{m_{l'}^2}{m_{l}^2} \right) \left. \right]   <
\frac{G_F}{\sqrt{2}} \frac{\alpha \sin^2 \theta_c \cos^2 \theta_c}{4
\pi \sin^2 \theta_W} \frac{m_c^2}{m_W^2} \label{54}
\eeq
where $l$ and $l'$  are leptons, and $\theta_c$ is the Cabbibbo angle.
If we neglect the second term on the left hand side, this
gives bounds on all vector leptoquarks coupling down-type quarks to
massive (= charged)  leptons $l$:
\beq
\lambda^{l1} \lambda^{l2} < .1 \left(
\frac{m_{lq}}{100 {\rm ~GeV}} \right)^2
\left( \frac{1 {\rm ~GeV}}{m_{l}} \right)
\eeq
We list the bounds for couplings to muons and taus in the tables.
Neglecting the term corresponding to the exchange of different flavoured
leptons
($l$ and $l'$) in the box should be an acceptable approximation, because
it would  be surprising if it cancelled against the first term,
and including it makes the bounds more complicated.

This analysis  has been for leptoquarks with left- or right-handed
couplings. One might hope that for $V_o^{\mu}$ and $V_{1/2}^{\mu}$, who
have both, one could escape the GIM-type suppression by having left-
and right-handed  couplings at either end of the lepton propagator.
This would give $\sum_l \lambda_L^{ld} \lambda_R^{ls} \neq 0$. However, the
lepton chiralities must be flipped in the propagators to do this,
giving a contribution to $\Delta m$ proportional to
$\lambda_L^2 \lambda_R^2m_l^2/m_{lq}^4$, so we will neglect these mixed bounds
(no
stronger; more difficult to calculate).

There has been no observed mixing between the $D^o$ and its
CP-conjugate the $\bar{D}^o$. The present upper bound on the mass
difference ($\Delta m_D < 1.3 \times 10^{-13}$ GeV) is considerably
larger than the Standard Model prediction, but can still be used to
constrain leptoquark couplings if we approximate the matrix element
as in (\ref{K1}) (with $f_D \sim .2$ GeV) . This gives
\beq
\frac{1}{32 \pi^2} | \lambda^{lu} \lambda^{lc}|^2 \frac{m_l^2}
{m_{lq}^4} < \frac{3
\Delta m_D}{2 f_D^2 m_D}
\eeq
implying
\beq
| \lambda^{lu} \lambda^{lc}| < .3
\left( \frac{m_{lq}}{100 {\rm ~GeV}} \right)^2
\left( \frac{1 {\rm ~GeV}}{m_{l}} \right)
\eeq
A similiar argument for the observed $B^o$-$\bar{B}^o$ mass
difference $\Delta m_B = 3.6 \times 10^{-13}$ GeV gives (using $f_B
\sim .2$ GeV)
\beq
| \lambda^{ld} \lambda^{lb}| < .3
 \left( \frac{m_{lq}}{100 {\rm ~GeV}} \right)^2
\left( \frac{1 {\rm ~GeV}}{m_{l}} \right)
\eeq
The limits for muons and taus ($l = 2,3$ on the internal lines)  are
listed in the tables.

We now briefly consider non-gauge leptoquarks.
The second term of the massive vector propagator:
\beq
\frac{i}{k^2 - m_{lq}^2} ( - g_{\mu \nu} + \frac{k_{\mu} k_{\nu}}{
m_{lq}^2}) \label{gp}
\eeq
is a nuisance in calculations, and the box diagrams we just discussed
were done in a gauge where it is absent. If the vector leptoquark is
not a gauge boson, we do not have this option, and  the loop momentum
integration over the $k^{\mu} k^{\nu}/m_{lq}^2$ terms diverges.
However, the new physics that produced the vector leptoquark
(compositeness?) should appear at some energy scale $\Lambda$, which
can be used as a cutoff for the divergent integral. The non-gauge
vector leptoquark contribution to $\Delta m$ is therefore
model-dependent. We can nonetheless get conservative  bounds on the
couplings  by only considering the contribution to the box amplitude
from the $g_{\mu \nu}$ terms. This is
\begin{eqnarray}
\Delta m_M =  \sum_{ij} (\lambda^{iq'} \lambda^{iq})( \lambda^{jq}
\lambda^{jq'}) \int \frac{d^4k}{(2\pi)^4}
\frac{k^{\mu} k^{\nu}}{(k^2 - m_{lq}^2)^2 (k^2 - m_i^2)(k^2 -
m_j^2)} \times  \nonumber \\
<M|(\bar{q}\gamma^{\alpha} \gamma_{\mu} \gamma^{\beta}
 P q')(\bar{q}\gamma_{\beta} \gamma_{\nu} \gamma_{\alpha} P
q')|\bar{M}>  \label{ngv}
\end{eqnarray}
where $M$  is a $K$, $D$ or $B$, $q$ and $q'$ are the meson
constituent quarks, $P$ is $L$ or $R$,  and $i$ and $j$ are lepton indices.
If there are no cancellations in the coupling constant sums (no
GIM-type suppression), the lepton masses can be neglected and the
integral approximated as $g^{\mu \nu}[64 \pi^2 m_{lq}^2]^{-1}$. Taking $i=j$,
and approximating the matrix element as in (\ref{K1}), we get
\beq
|\lambda^{iq} \lambda^{iq'} |^2 < \frac{48 \pi^2 m_{lq}^2}
{f_M^2 m_M}\Delta m_M
\eeq
for non-gauge vector leptoquarks. This gives the numerical
bounds
\begin{eqnarray}
\lambda^{ld} \lambda^{ls} & < &  6 \times 10^{-4}
 \left( \frac{m_{lq}}{100{\rm GeV}}\right) ~~(K \bar{K}) \\
\lambda^{lu} \lambda^{lc} & < & 2 \times 10^{-3}
 \left( \frac{m_{lq}}{100{\rm GeV}}\right) ~~(D \bar{D})\\
\lambda^{ld} \lambda^{lb} & < & 2 \times 10^{-3}
 \left( \frac{m_{lq}}{100{\rm GeV}}\right) ~~(B \bar{B})
\end{eqnarray}
where $l$ is an arbitrary lepton flavour index.
 We do not list the bounds in the
tables because they only apply if there is no ``GIM'' suppression.

\subsection{Scalar leptoquarks}

As we mentioned, the constraints on scalars are more interesting. The
box diagrams for a scalar leptoquark (see figure 2) with couplings to
leptons of only one chirality (we neglect diagrams proportional to
$(\lambda_{L}\lambda_{R})^2$) give a contribution to
$\Delta m_M$ of
\begin{eqnarray}
\Delta m_M =  \sum_{ij} (\lambda^{iq'} \lambda^{iq})
( \lambda^{jq}\lambda^{jq'}) \int \frac{d^4k}{(2\pi)^4}
\frac{k^{\mu} k^{\nu}}{(k^2 - m_{lq}^2)^2 (k^2 - m_i^2)(k^2 -
m_j^2)} \times \nonumber \\
<M|(\bar{q} \gamma_{\nu} P q')(\bar{q} \gamma_{\mu} P
q')|\bar{M}> \label{slq}
\end{eqnarray}
with the same notation as in (\ref{ngv}). This is  the same integral
and expectation value as for vectors, but the scalar coupling
constant matrix is not required to be unitary,    so the integral can
be approximated as $g^{\mu \nu}[64 \pi^2 m_{lq}^2]^{-1}$, as in the non-gauge
vector case. If we take take $i=j$ and assume there are no
cancellations between the different lepton contributions to $\Delta
m_M$, then we get the constraint
\beq
|\lambda^{iq} \lambda^{iq'} |^2 < \frac{196 \pi^2 m_{lq}^2 \Delta m_M}
{f_M^2 m_M}
\eeq
where we have used equation (\ref{K1}) to approximate the matrix
element. Note that these constraints are independant of the lepton
mass, so apply equally to neutrinos and all generations of charged
leptons. We use $f_D \simeq .2 $ GeV $\sim f_B$ to get the numbers in
the tables.

\section{Muon Physics}

Muon family number violating processes provide some  of the strongest
constraints on leptoquarks from leptonic physics. Bounds on
scalar leptoquarks from muon conversion on nuclei, $\mu \rightarrow
e \bar{e} e$, and $\mu \rightarrow e \gamma$ have been considered in
\cite{S2,MSW,BKZ,CEEGN,D+H}.

There is a very strong upper bound on the lepton flavour-changing
rate for a muon to turn into an electron when scattered off a
nucleus. The experimental limit \cite{Bry} on muon conversion on
titanium is
\beq
\frac{\Gamma( \mu {\rm Ti} \rightarrow e {\rm Ti})}{\Gamma( \mu {\rm
Ti} \rightarrow \nu {\rm Ti'})} < 4.6 \times 10^{-12}~~.
\eeq
If we neglect the difference in kinematics of the neutrino and the
electron, this implies
\beq
\left| \frac{e_L^2}{m_{lq}^2} <e{\rm Ti}|(\bar{\mu} \gamma^{\mu} P
e)(\bar{q} \gamma_{\mu} P q)|\mu {\rm Ti}> \right|^2 <  4.6 \times
10^{-12} \left| \frac{4 G_F}{\sqrt{2}}<\nu {\rm Ti}|(\bar{\mu}
\gamma^{\mu} L \nu)(\bar{u} \gamma_{\mu} L d)|\mu {\rm Ti}> \right|^2
  \eeq
where $P$ is a chiral projection operator. The matrix elements are not
necessarily the same, because the isospin structure could be different,
and $\mu$-$e$ conversion could be a coherent process \cite{S}. The
ratio of the matrix elements was calculated in \cite{S,C+H} for
a coherent $E$-$\mu$ conversion interaction, and the $N \mu \rightarrow
N e$ matrix element was found to be larger than that for  $N \mu \rightarrow
N' \nu$. However, the interaction can only be coherent if it is
vector, and one can arrange the couplings of some of the leptoquarks
such that they are only axial vector. We will therefore assume that
the matrix elements are comparable and require
\beq
\frac{e_L^4}{m_{lq}^4} < 4 \times 10^{-11}  G_F^2
\eeq
which applies to all the leptoquark interactions because they all
couple  $u$ or $d$ quarks to charged leptons.
As usual, the precise limits are in the tables.

\subsection{ $\mu \rightarrow ee\bar{e}$}

Leptoquarks could allow the decay of a muon to $ee\bar{e}$ via box
diagrams (see figures 1 and 2), or through the effective coupling of
a muon to an electron and a $Z$ (or a $\gamma$*) which subsequently decays to
$e\bar{e}$. The box amplitudes are similiar  to those for
meson-anti-meson mixing, with quarks and leptons interchanged. The
effective $Z \mu e$ vertex is induced by leptoquark triangle
diagrams, is divergent, and is non-trivial to calculate. By analogy
with the Standard Model \cite{G+L}, we expect  the triangle amplitude
to be larger than the box by a factor of $\ln(m_{lq}^2/m_q^2)$; we
will nonetheless only discuss the box diagrams because they are
easier to compute; this will give us conservative limits.

Scalar and vector leptoquark boxes will induce  effective four-lepton
vertices
\beq
C_{LL} (\bar{\mu} \gamma^{\mu} L e) (\bar{e} \gamma_{\mu} L e)
{}~,~~C_{RR} (\bar{\mu} \gamma^{\mu} R e) (\bar{e} \gamma_{\mu} R e)
\label{m1}
\eeq
In the Standard Model, the muon decays principally (BR$ \simeq
100\%$) to $e \bar{\nu}_e \nu_{\mu}$ via a $W$- induced four-lepton
vertex similiar to (\ref{m1}):
\beq
\frac{4 G_F}{\sqrt{2}} (\bar{\mu} \gamma^{\mu} L \nu) (\bar{\nu}
\gamma_{\mu} L e)
\eeq
If we neglect the electron mass and the difference in the matrix element
arising from having identical fermions in the final state, we can
constrain the effective couplings $C_{LL}, C_{RR}$ by
requiring
\beq
(C_{LL,RR})^2 < 8 G_F^2 \frac{BR( \mu \rightarrow 3 e)}{BR( \mu
\rightarrow \nu \bar{\nu} e)}   \label{m2}
\eeq
where $BR( \mu \rightarrow 3 e)$ of course means the experimental
upper bound \cite{PDB} of $ 10^{-12}$.

Had we defined an effective
coupling like $C_{LL}$ for the meson-anti-meson mixing, we would have had
$\Delta M/2 = C_{LL} \times$ [hadronic matrix element], so we can estimate
$C_{L
L}$ and $C_{RR}$ from the meson-anti-meson  mixing equations.
{}From (\ref{K0}) we have for gauge vector leptoquarks
\beq
C_{LL}, C_{RR} \simeq \frac{3}{32 \pi^2} (\lambda^{1q})^3 \lambda^{2q}
\frac{m_q^2}{m_{lq}^4} \label{qlu}
\eeq
where we have assumed the same quark flavour on both internal fermion lines,
multiplied by three for colour, and neglected the second term of
(\ref{54}) for simplicity.
This does not give very
interesting constraints on the gauge leptoquarks because the
amplitude is ``GIM'' suppressed, giving a decay rate proportional to
$\lambda^8 m_q^4/m_{lq}^8$. However, it does give bounds on couplings of
electrons and muons to quarks of all flavours (except $t$). We have, from
(\ref{m2})
\beq
\sqrt{ \lambda^{2q} (\lambda^{1q})^3} < .6 \left(
 \frac{m_{lq}}{100 {\rm ~GeV}}
\right)^2  \left( \frac{1 {\rm ~GeV}}{m_{q}} \right) ~~~
({\rm gauge\ vectors})  \label{m3}
\eeq
where $q$ is a quark flavour other than top.

 For non-gauge vectors we have, in
the same approximation used for meson-anti-meson mixing:
\beq
C_{LL}, C_{RR} = \frac{3(\lambda^{1q})^3 \lambda^{2q}}{32 \pi^2 m_{lq}^2}
\label{5.3'}
\eeq
This gives a more interesting bound than (\ref{m3}):
\beq
\sqrt{ \lambda^{2q} (\lambda^{1q})^3} < 6 \times  10^{-3} \left(
\frac{m_{lq}}{100 {\rm ~GeV}} \right) ~~({\rm non-gauge\ vectors})
\label{5.9b}
\eeq
where again $q \neq t$.
We do not list the non-gauge results in the tables.

 Scalar leptoquarks give
\beq
C_{LL}, C_{RR} = \frac{3(\lambda^{1q})^3 \lambda^{2q}}{128 \pi^2 m_{lq}^2}
\label{5.3}
\eeq
for light quarks, which implies ($q \neq t$)
\beq
\sqrt{ \lambda^{2q} (\lambda^{1q})^3} <  10^{-2} \left(
\frac{m_{lq}}{100 {\rm ~GeV}} \right) ~~({\rm scalars}).\label{5.9}
\eeq

These approximations for $C_{LL}$ and $C_{RR}$ only apply if the internal
line fermions are light.  For meson-anti-meson mixing induced by
leptoquarks, this is a good approximation because the internal
line fermions are leptons. However, one
 could in principle get constraints on leptoquark
 couplings to the top quark from $\mu \rightarrow e \bar{e} e$ box
diagrams involving top quarks and leptoquarks. The exact
expression for the intergral in (\ref{ngv}) and (\ref{slq}) is \cite{C+L}
\beq
\frac{1}{64 \pi^2 m_{lq}^2} \frac{1}{x_t - x_q} \left[ \right.
\frac{1}{1 - x_q} - \frac{1}{1 - x_t} + \frac{x_q^2}{(1 - x_q)^2}
\ln x_q^2 -  \frac{x_t^2}{(1 - x_t)^2}
\ln x_t^2 \left. \right]
\eeq
where
$ x_t = m_t^2/m_{lq}^2$, $ x_q = m_q^2/m_{lq}^2$, and
$q$ is any up-type quark.  Any bounds will clearly depend sensitively on
 the top and leptoquark masses, so we neglect them here.

\subsection{$\mu \rightarrow e \gamma$}

The lepton flavour violating decay $\mu \rightarrow e \gamma$ could
be mediated by leptoquark triangle diagrams (see figures 3 and 4).
This decay will again be ``GIM" suppressed for gauge leptoquarks, but in this
case will be of order $\lambda^4 m_q^4/m_{lq}^8$ rather than
$O(\lambda^8 m_q^4/m_{lq}^8)$
as in the $\mu \rightarrow e\bar{e}e$ case.  There are also fewer powers
of $\lambda$ in the scalar and non-gauge vector rates.

The decay rate of an initial fermion $f_i$ into a photon and a final fermion
$f_f$ is
\cite{RND}
\beq
\Gamma (f_i \rightarrow f_f \gamma) = \frac{m_i}{8 \pi} \left( 1 -
\frac{m_f}{m_i} \right)^2 \left( 1 - (\frac{m_f}{m_i})^2 \right)
\left[ F^V(0)^2 + F^A(0)^2 \right]  \label{m5}
\eeq
where  $F^V$  and  $F^A$ are the vector and axial vector coefficients
of the magnetic moment term in the matrix element:
\beq
{\cal M}^{\mu} [f_i(p_i) \rightarrow f_f(p_f) + \gamma (q)] =
-i\bar{u}_f(p_f) \left\{ \ldots + \frac{ \sigma^{\mu \nu}
q_{\nu}}{m_f + m_i} [ F^V(q^2) + F^A(q^2) \gamma_5] + \ldots \right\}
u_i(p_i) \label{m5a}
\eeq
and $\sigma^{\mu \nu} = \frac{-i}{2} [\gamma^{\mu}, \gamma^{\nu}]$.

The $\lambda_L^2$ and $\lambda_R^2$ terms in the scalar leptoquark
contribution to $(g-2)_e$ have been calculated
in detail in \cite{DKST,Lev,D+H}. Djouadi, Kohler,
 Spira and Tuta \cite{DKST}, and Morris \cite{Mor}
also calculate the contribution
proportional to  $\lambda_L \lambda_R$.
These should be closely related to $F_V$ and $F_A$;  if one set
$m_i = m_f = m_e$ in the final result for $F_V$, one ought to
get $e (g-2)_e = F_V$. Unfortunately, in this limit our approximate
result is smaller than the exact one in \cite{DKST} by a factor of 1/3, and
has a different coefficient in front of the leptoquark charge. In
spite of this disturbing discrepancy, we will use our formula
because it gives weaker  bounds on the couplings, and
because it is three times the result in \cite{Lev} (three for
colour), whose results can be checked against Standard Model
calculations.(Davies and He \cite{D+H} seem to multiply Leveille's
result by 9 in their published paper, which would agree with \cite{DKST}.
This is apparently a typographical error, and should have been a three
\cite{HePC}.) Note that  we have also
neglected QCD corrections to the diagrams.

If we assume $m_q \ll m_{lq}$ (this assumption was not made in
\cite{DKST,Lev}, so their result is applicable to all quarks, in particular the
top), we can approximate
\beq
\frac{|F^V_+|}{m_i + m_f} = \frac{|F^A_-|}{m_i + m_f}  \simeq
\frac{\lambda_L^2}{(8 \pi)^2}
\frac{m_{i} \pm m_f}{m_{lq}^2} e(\frac{Q_{lq}}{2} - Q_q)  ~~~~({\rm scalars})
\label{m6b}
\eeq
for couplings of the same chirality at either end of the scalar leptoquark
propagator. (The same bound applies to  $\lambda_R^2$.)
 $Q_q (Q_{lq}$) is the quark (leptoquark) electric charge coming
out of the incident muon vertex, ``$e$'' is
the electromagnetic coupling constant, and $\lambda^{lq}$ is an
orthogonal matrix in our approximation that all coupling constants
are real. The $+$ and $-$ on the RHS of (\ref{m6b}) are respectively for
$F^V$ and $F^A$ (hence the subscipts on $F^V$ and $F^A$).

The $\lambda_L \lambda_R$ contribution (from scalar leptoquarks) can
be larger than the $\lambda_L^2$, $\lambda_R^2$ terms, because it is
proportional to $(m_i m_q/m_{lq}^2) \ln[m_{lq}^2/m_q^2]$. We will nonetheless
neglect it  because it only  applies to those leptoquarks which have couplings
of both chiralities. These terms are present in the calculation of
\cite{Mor,DKST}.

Requiring that radiative muon decay, as calculated from (\ref{m5}) and
(\ref{m6b}) take place more slowly than the present
experimental upper bound $\Gamma(\mu \rightarrow e \gamma) < 1.5 \times
10^{-29}$ GeV, gives
\beq
\lambda_L^{2q} \lambda_L^{1q}, \lambda_R^{2q} \lambda_R^{1q} < 8 \times
10^{-5} \left( \frac{m_{lq}}{100 {\rm GeV}} \right)^2 \label{fea}~~~~
({\rm scalars})
\eeq
where $q$ is a quark other than the top,
and we have assumed here (but not in the tables) that $\frac{Q_{lq}}{2}
 - Q_{q} \simeq 1$. For the $SU(2)$ doublet and triplet leptoquarks,
this formula is an  underestimate, because it assumes that
only one leptoquark is being exchanged (rather than two or
three of different charges). For simplicity, we estimate the rate as
being due to the member of the mutiplet that gives the largest
$(\frac{Q_{lq}}{2} - Q_q) $. For $\tilde{S}_{1/2}$,
$(\frac{Q_{lq}}{2} - Q_q) = 0$ (remember that $Q_q$ and $Q_{lq}$ are defined
as the charges coming out of the first vertex), so we have no
bound on the tables. There will certainly be a contribution
to $F_V$ from $\tilde{S}_{1/2}$ leptoquarks, but it will be suppressed with
respect to (\ref{m6b}) by powers of $m_q/m_{lq}$ or $m_{\mu}/m_{lq}$,
so we neglect this. (It is also possible that ou
approximate formula is wrong, and $F^V \sim a Q_{lq} + Q_q$ with
$a \neq -1/2$, in which case the bound would be of order (\ref{fea}).)

The two  triangle
diagrams for gauge vector leptoquarks contribute approximately
\beq
\frac{|F^V_+|}{m_i + m_f} = \frac{|F^A_-|}{m_i + m_f}
\simeq  \frac{\lambda_L^{fq} \lambda_L^{iq} e}{32 \pi^2}
\frac{ (m_i \pm m_f)}{m_{lq}^2} \times \nonumber
\eeq
\beq
\left(Q_q\left[2 - \frac{m_q^2}{m_{lq}^2}
(\frac{7}{2} + 3 \ln \frac{m_q^2}{m_{lq}^2})  \right]
   +  Q_{lq}\left[\frac{5}{2} - \frac{13}{4}\frac{m_q^2}{m_{lq}^2} \right]
\right)  ~~~~({\rm gauge\ vectors})
\label{m6}
\eeq
where  we have neglected
a number of (presumably smaller) terms, and the same formula
would apply to couplings $\lambda_R^2$.  The term of order
 $[\lambda \lambda^T]_{if}/m_{lq}^2$
 vanishes for $i \neq f$ due to the orthogonality of $\lambda^{lq}$.

If we set $ m_i = m_f = m_{\mu}$, we can check this
expression by comparing it to Leveille's \cite{Lev} formula for
the massive gauge boson contribution to $g-2$. Requiring
\beq
\Gamma(\mu \rightarrow  e \gamma) \simeq \frac{\alpha }{(4 \pi)^4}
\frac{(\lambda_L^{2q} \lambda_L^{1q})^2 m_q^4}{m_{lq}^8} \,
m_{\mu}^5 \left[ Q_q\left(\frac{7}{2} + 3 \ln \frac{m_q^2}{m_{lq}^2})  \right)
   +   \frac{13}{4} Q_{lq} \right]^2  < 1.5 \times 10^{-29} {\rm ~GeV}
\label{m7}
\eeq
gives
\beq
\lambda_L^{2q} \lambda_L^{1q}, \lambda_R^{2q} \lambda_R^{1q} < \frac{.2}
{\left[ Q_q\left(\frac{7}{2} + 3 \ln (\frac{m_q^2}{m_{lq}^2})  \right)
   +   \frac{13}{4} Q_{lq} \right]} \left(
\frac{m_{lq}}{100 {\rm ~GeV}} \right)^4 \left( \frac{1 {\rm
{}~GeV}}{m_{q}}\right)^2 ~~~~({\rm gauge\ vectors})
\eeq
for gauge vector couplings. This
 applies to all quark flavours except  top. Due to the
number of approximations in this calculation, this constraint can
only be considered a rough estimate.

It would not be difficult to modify the Standard Model amplitude for
$b \rightarrow s \gamma$ to descibe a leptoquark and a quark of
arbitrary charge in the loop. Such a ``modified Standard Model''
expression could be used to constrain gauge vector leptoquark
couplings to the top quark. However, as in the case of meson
anti-meson mixing, the formula is fairly complicated, and does not
provide clear constraints on the coupling constant when both the top
and the leptoquark masses are unknown.

Some of the vector leptoquarks ($V_o^{\mu}, V_{1/2}^{\mu}$) have two different
interactions with Standard Model fermions. The two couplings
have opposite chiral projection operators. We argued that
for the box diagrams, the bounds on the product $\lambda_L\lambda_R$ would
be of the same magnitude as the ``GIM" suppressed bounds on $\lambda_L
\lambda_L$
or $\lambda_R\lambda_R$ because the fermion mass must appear in the amplitude
to flip the chirality between the gauge boson vertices. Since
the effective coupling for the four-fermion vertex has dimension mass$^{-2}$,
it must be of order $\lambda_L^2 \lambda_R^2
m_f^2/m_{lq}^4$, where $m_f$ is the
mass of the internal line fermions. This is roughly the
same as the ``GIM'' suppressed amplitude constraining $\lambda_L^2$ or
$\lambda_R^2$. However, this is not true for
 the triangle graphs, where the bounds on
$\lambda_L\lambda_R$ are much better than those on $\lambda_L^2$ or
$\lambda_R^2$: the coefficients $F_V$ and $F_A$ are dimensionless, so
putting a quark mass in the numerator to flip the chirality
simply changes the leading contribution to $F_V$ and $F_A$ from
$\lambda_L^2  m_{\mu}^2/m_{lq}^2$ to $\lambda_L \lambda_R m_{\mu}
m_q/m_{lq}^2$.
 Since there is no reason to expect $\sum_q \lambda_L^{\mu q}
\lambda_R^{e q} = 0$ for vector leptoquarks, we have \cite{MS}
\beq
 \frac{|F^V_+|}{m_i + m_f} = \frac{|F^A_-|}{m_i + m_f}  \simeq
 \frac{3  e }{16 \pi^2}
\frac{m_q}{m_{lq}^2} (\lambda_L^{iq} \lambda_R^{fq} \pm
\lambda_L^{fq} \lambda_R^{iq}) (Q_{lq} - Q_{q})~~~~({\rm vectors})
\label{ope}
\eeq
where $e$ is the electromagnetic coupling constant,  and $i$ and $f$
respectively label the initial and
final leptons. Taking $ (Q_{lq} - Q_{q}) \sim 1$ (we do not assume
this in the tables), this gives
\beq
 \lambda_L^{2q}  \lambda_R^{1q},  \lambda_R^{2q}  \lambda_L^{1q}
 <  10^{-6}  \left(
\frac{m_{lq}}{100 {\rm ~GeV}} \right)^2 \left(
\frac{1 {\rm ~GeV}}{m_{q}}\right) ~~~~({\rm vectors})
\eeq
from the upper bound on the decay $\mu \rightarrow e \gamma$.

For non-gauge vectors, the leptoquark propagator is again of the
form (\ref{gp}), and the second term will give cutoff dependant
contributions to the decay rate. If the leptoquark mass is well
below the scale at which new physics appears, these terms could be large.
 We will nonetheless neglect them, as we did for the boxes. This
should make our bounds conservative. So using only the $g_{\mu \nu}$ part of
the vector propagator, and assuming that there are no cancellations
between triangles containing different quark flavours, we get
the leading order terms of (\ref{m6}):
\beq
 \frac{|F^V_+|}{m_i + m_f} = \frac{|F^A_-|}{m_i + m_f}
 \simeq  \frac{\lambda_L^{1q} \lambda_L^{2q} e }{32 \pi^2} \left(
\frac{m_i \pm m_f}{m_{lq}^2} \right) (2Q_q + \frac{5}{2} Q_{lq})  ~~~~({\rm
non-gauge\ vectors}). \label{ngv2}
\eeq
This gives
\beq
\lambda^{2q} \lambda^{1q} <  2 \times 10^{-5}  \left(
\frac{m_{lq}}{100 {\rm ~GeV}} \right)^2  ~~~~({\rm non-gauge\ vectors})
\label{ngvnum}
\eeq
where we have again taken $ (2 Q_q + \frac{5}{2} Q_{lq}) \approx 2$.

\section{$\tau$ decays}

\subsection{Semi-leptonic decays}

 \paragraph{}
The tau is the only lepton heavy enough to decay to a meson and
another lepton. The Standard Model lepton family number conservation
only allows decays to $\nu_{\tau} + X$, and $\tau \rightarrow \pi \nu$
is observed at the expected rate. The leptoquark  couplings
$\lambda_{LS_o}, \lambda_{LS_1}, \lambda_{LV_o}$ and $\lambda_{LV_1}$,
which couple a tau,
a neutrino, and an up and down type quark, must therefore satisfy
\beq
\frac{e^2}{m_{lq}^2} < \frac{4 G_F}{\sqrt{2}}~~.
\eeq
Leptoquarks could also
mediate $\tau \rightarrow e M, \mu  M$, where $M$ is a neutral
 meson lighter than the
tau.
 If we assume that
 \beq
 <0|\bar{u^i} \gamma^{\mu} \gamma^5 u^i|M^o> \simeq  <0|\bar{d^i}
\gamma^{\mu} \gamma^5 d^i|M^o>
 \simeq <0|\bar{u^i} \gamma^{\mu} \gamma^5 d^i|M^->
 \eeq
(for instance, $<0|\bar{u} \gamma^{\mu} \gamma_5 u|\pi^o> \sim
<0|\bar{d} \gamma^{\mu} \gamma_5 u|\pi^+>$),
and neglect the mass of the decay
lepton, then we do not need to calculate the tau decay rate to a
lepton ($\ell$) and a meson($M$). For a four-vertex of the form
\beq
  \frac{e^{3k} e^{ij}}{m_{lq}^2} (\bar{\tau} \gamma^{\mu} P
\ell^i)(\bar{q}^j \gamma_{\mu} P q^k)
\eeq
we can simply require
\beq
 \left( \frac{e^{3k} e^{ij}}{m_{lq}^2} \right)^2 < 8 G_F^2 |V|^2
\frac{BR(\tau^- \rightarrow M^o \ell^-)}{BR(\tau^- \rightarrow M^-
\nu)}
 \eeq
 where $V$ is the appropriate CKM matrix element for the decay
$\tau^- \rightarrow M^- \nu$, $k$ and $j$ label the
 flavours of the constituent quarks  in $M$,
and $BR(\tau^- \rightarrow M^o \ell^-)$
is the upper bound on this process \cite{ARGUS}. Using $\ell^i = e, \mu$ and
$M^o
= \pi^o, K^o$, we get the constraints listed in the tables. Note that
unlike the pseudoscalar meson decays, there is no chiral suppression
so we would not get better bounds on the couplings of the effective
pseudoscalar vertices. We therefore do not compute constraints on the
product of left- and right-handed couplings.

 \subsection{Leptonic decays}

 \paragraph{}
 Leptoquarks will contribute to the decays $\tau \rightarrow \ell
\bar{\ell} \ell$ and $\tau \rightarrow \ell \gamma$ ($\ell = \mu,e$)
in the same way as they did for muons.  The bounds however, will be
weaker, because the rare decays of the tau are not as tightly
constrained as those of the muon.  Using  equations (\ref{5.3}),
(\ref{5.3'}) and
BR$(\tau \rightarrow 3 \ell) \lappeq 1.5 \times 10^{-5}$ \cite{ARGUS},
we find that for scalar and non-gauge vector leptoquarks
 \beq
 \lambda_L^{3q} (\lambda_L^{lq})^3,
\lambda_R^{3q} (\lambda_R^{lq})^3 < .2 \left( \frac{m_{lq}}{100 {\rm ~GeV}}
\right)^2  ~~~({\rm non-gauge\ vectors})  \label{cc}
 \eeq
 \beq
 \lambda_L^{3q} (\lambda_L^{lq})^3,
\lambda_R^{3q} (\lambda_R^{lq})^3 < .3 \left( \frac{m_{lq}}{100 {\rm ~GeV}}
\right)^2  ~~~({\rm scalar})  \label{cc'}
 \eeq
 where $q$ is any quark flavour other than top, and $\ell$ can be
$\mu$ and/or $e$.
The bounds on gauge vector leptoquarks are to weak to be
meaningful.

 The experimental upper bounds on the decays $\tau \rightarrow \ell
\gamma$ are $\Gamma(\tau \rightarrow \mu \gamma) < 9.2 \times
10^{-18}$ GeV \cite{CLEO} and $\Gamma(\tau \rightarrow e \gamma) < 2.6 \times
10^{-16}$ GeV \cite{ARGUS}. The radiative muon decay bounds can be
easily modified to constrain leptoquarks coupling to the $\tau$. The
constraints on gauge vectors with couplings of one chirality are again to
weak to be worth listing, but equations  (\ref{m5}) and  (\ref{ope}) give
bounds on vector couplings of both chiralities ($\lambda_L \lambda_R$),
which are  in
the tables. One can also get bounds on non-gauge vectors from (\ref{ngv2})
and the upper bounds on $\tau \rightarrow \mu \gamma$ and
$\tau \rightarrow e \gamma$. This gives
\beq
 \lambda_L^{3q} \lambda_L^{2q},
\lambda_R^{3q} \lambda_R^{2q} < \frac{10^{-2}}{Q_q + \frac{5}{4} Q_{lq}}
 \left( \frac{m_{lq}}{100 {\rm ~GeV}} \right)^2~~~({\rm non-gauge\ vectors})
\eeq
\beq
 \lambda_L^{3q} \lambda_L^{1q},
\lambda_R^{3q} \lambda_R^{1q} < \frac{.07}{Q_q + \frac{5}{4} Q_{lq}}
 \left( \frac{m_{lq}}{100 {\rm ~GeV}} \right)^2~~~({\rm non-gauge\ vectors})
\eeq
 Equations (\ref{m5})and  (\ref{m6b}) can be
used to constrain scalar leptoquarks contributing to radiative tau decays.
We list the bounds in the tables.

\section{Other Constraints}

\subsection{Quark-lepton universality}

\paragraph{}
As was pointed out in \cite{BRW}, leptoquarks
would contribute at tree level to neutron $\beta$-decay,
but only via box diagrams to $\mu \rightarrow e \nu \bar{\nu}$. We
know from (\ref{qlu}), (\ref{5.3'}) and (\ref{5.3})  that the effective
leptoquark
induced coupling constant for the 4-lepton vertex is much smaller
than that for the 2-quark 2-lepton vertex:
 \beq
 C_{LL,RR} \ll \frac{\lambda^2}{m_{lq}^2}
\eeq
 We can therefore neglect the leptoquark contribution to muon decay,
and obtain constraints on the $ude\nu$ couplings from requiring that
the Fermi constant in $\beta$ decay not differ signifcantly from the
muon decay measurement.  If the effective Fermi constant for $\beta$
decay $\equiv G_{\beta}$ was
 \beq
 \frac{4 G_{\beta}}{\sqrt{2}}  = \frac{4G_F}{\sqrt{2}} +
\frac{ e_L^2}{m_{lq}^2}
\eeq
 then the rate  for $n \rightarrow p e \bar{\nu}$ would be increased
by a multiplicative factor
 \beq
 1 + \frac{e_L^2}{\sqrt{2} G_F m_{lq}^2} + \frac{e_L^4}{8 G_F^2 m_{lq}^4}
\eeq
The CKM matrix element $V_{ud}$ is determined to be
0.9744 $\pm .0010$ \cite{PDB} by comparing $G_{\beta}$ to $G_F$, and
this value is consistent with unitarity.  We can therfore require that
the leptoquarks contribute less than the error in $V_{ud}$:
\beq
 \frac{\sqrt{2}e_L^2}{4 m_{lq}^2} <  \frac{\delta V_{ud}}{V_{ud}} G_F
\eeq
where we take $\delta V_{ud} = 0.9754 -$  [the minimum value
given in the Particle Data Book] \cite{PDB}. This gives
\beq
e_L^2 < 10^{-3} \left( \frac{m_{lq}}{100 {\rm GeV}} \right)^2
\eeq
 for $ude\nu$ axial vector couplings. We neglect the contributions
from effective scalar $\pm$ pseudoscalar quark matrix elements ($\sim
\lambda_L\lambda_R$), because they are strongly constrained
 by the ratio $R_{\pi}$. (They  have also been carefully considered
in \cite{Her}.)

\subsection{$g$ - 2}

Since the magnetic moment of the electron and the muon are two of
the most accurately measured quantities in physics, one always
hopes to get good constraints on new particles from them.
Constraints on leptoquarks from $g-2$ have been calculated in
\cite{D+H} and \cite{DKST}. It is easy to see
from equation (\ref{m5a}) that
\beq
\frac{(g-2)_e}{2} = \frac{m_e}{e} \frac{F_V}{m_i + m_f}~.
\eeq
with $m_i = m_f = m_e$.
Using $F_V/(m_i + m_f)$ from equations (\ref{m6b}),
(\ref{m6}), (\ref{ope}), and (\ref{ngv2}), we can constrain some
of the couplings by requiring that the leptoquark contribution to
$g-2$ for the muon and the electron be less than the difference between the
theoretical predictions and the experimental measurements. Although
$[g-2]_e$ is more accurately measured than $[g-2]_{\mu}$ ($\Delta[g-2]_e/2
= 6 \times 10^{-10}, \Delta[g-2]_{\mu}/2 = 2 \times 10^{-8}$), the
leptoquark contributions are proportional to the lepton mass, or mass squared,
so the relative smallness of $m_e$ suppresses most leptoquark
contributions to $[g-2]_e$. As can be seen from the tables, these bounds
are considerably weaker than those from radiative decays, but constrain
different combinations of generation indices. We have neglected
couplings to the top, because we assumed $m_q \ll m_{lq}$ in
calculating $F_V$. As previously mentioned, this approximation was not made by
Davies and He \cite{D+H}, or
Djouadi, Kohler, Spira and Tutas \cite{DKST}, who give a full
analytic expression for  the scalar leptoquark contribution to
$g-2$, and a reference \cite{Spira} for the non-gauge vector
contribution where the divergent $k^{\mu} k^{\nu}/m_{lq}^2$ term
in the propagator is included.

{}From equation (\ref{m6}), we can get weak constraints  on the gauge vector
leptoquarks that contribute to $g-2$. Since the initial and final lepton
are of the same type in this diagram, there is no ``GIM'' suppression,
and the leading order terms in (\ref{m6}) give the bound
\beq
 \lambda_L^{2q} \lambda_L^{2q},
\lambda_R^{2q} \lambda_R^{2q} < \frac{ 2}{Q_q + \frac{5}{4} Q_{lq}}
 \left( \frac{m_{lq}}{100 {\rm ~GeV}} \right)^2 \label{nic}
\eeq
where $q$ is any quark other than top. The bounds on $\lambda_L \lambda_R$
for gauge vectors are stronger than this (for heavy quarks in the loop).
We get, from $(g-2)_e$ and $(g-2)_{\mu}$:
\beq
\lambda_R^{2q} \lambda_L^{2q} < \frac{.1}{Q_{lq} - Q_{q}}
 \left( \frac{m_{lq}}{100 {\rm ~GeV}} \right)^2 ~~~~
\eeq
\beq
  \lambda_L^{1q} \lambda_R^{1q} < \frac{ .6}{Q_{lq} - Q_{q}}
\left( \frac{m_{lq}}{100 {\rm ~GeV}} \right)^2 ~~~.
\eeq

The bounds on scalar leptoquark couplings of the same chirality
($\lambda_L^2, \lambda_R^2$) from $(g-2)_{\mu}$  are similiar to
(\ref{nic}). As in the vector case, we could probably get stronger bounds on
$\lambda_L \lambda_R$, at least for heavy quarks on the internal lines.
We neglect this, because there are better constraints from elsewhere
(see table 15), and bounds on the combination $\lambda_L \lambda_R$
are less interesting in the first place.

\subsection{Neutral current couplings}

\paragraph{}
The many different experimental determinations of $\sin^2 \theta_W$ seen to
agree quite well. Since leptoquarks mediate two-lepton, two-quark
interactions like the weak gauge bosons of the Standard Model, they
could contribute at tree level in some, but not all, of these experiments.
 There will
therefore be bounds on leptoquark couplings from requiring
that they not disrupt the agreement between various determinations
of $\sin^2 \theta_W$. These will be particularily interesting,
because they apply to leptoquarks coupling to first
generation fermions-- precisely the leptoquarks that could be seen
at HERA.

Leptoquarks should not have a significant effect
on the LEP measurements, because
LEP runs on the $Z$ mass. However, they
should contribute on an equal footing with the electroweak gauge bosons
in atomic parity violation experiments, and in deep inelastic scattering of
neutrinos off nucleons.
It turns out that the best bounds on the largest
number of leptoquark  coupling constants come from
atomic parity violation experiments on cesium\cite{PDB}, and
these constraints have been considered by Langacker \cite{Lang}. The
atomic weak charge $Q_W = -2 [C_{1u}(2Z + N) + C_{1d}(Z + 2N)]$
of the cesium atom is
measured to be $Q_W = -71.04 \pm 1.58 \pm .88$  (the second
error is theoretical). $C_{1u}$ and $C_{1d}$ are coefficients in the
effective four-fermion parity-violating Lagrangian
\beq
\frac{G}{\sqrt{2}} (\bar{l} \gamma^{\mu} \gamma_5 l)(C_{1u}\bar{u}\gamma_{\mu}
u + C_{1d} \bar{d} \gamma_{\mu} d)
\eeq
where at tree level  (see \cite{PDB} for the one-loop Standard Model
corrections) $C_{1q} = - I_3 + 2 Q_{em}\sin^2 \theta_W +\Delta C_{1q}$, and
$\Delta C_{1q}$ is the leptoquark contribution:
\beq
\Delta C_{1q} = \frac{e^2}{m_{lq}^2} \frac{\sqrt{2}}{4 G_F} ~.
\eeq
The Standard Model expectation for $Q_W$ is $-73.21 \pm.08 \pm .03$, so
requiring that the leptoquark contribution to $Q_W$ be less than the
difference between the experimental and Standard Model values, gives
\beq
|\Delta C_{1u}|=|\Delta C_{1d}| < .003 \label{apvbd1}
\eeq
for leptoquarks that induce $ \Delta C_{1u} = \Delta C_{1d}$, and
\beq
|\Delta C_{1u}| , |\Delta C_{1d}| < .006 \label{apvbd2}
\eeq
for leptoquarks where $\Delta C_{1u}= 0$ or $\Delta C_{1d}= 0$. These
bounds are slightly weaker than those in \cite{Lang}. They also
neglect the sign of the corrections, which is conservative,
but not altogether
reasonable because the theoretical prediction is one and half $\sigma$ to one
side of the experimental result. From (\ref{apvbd1}), (\ref{apvbd2})
and tables 3 and 4, one gets that for $\lambda_{LV_1}, \lambda_{RV_{1/2}},
 \lambda_{RS_{1/2}}$ and  $\lambda_{LS_1}$
\beq
e^2 < .001 \left(\frac{m_{lq}^2}{100 {\rm GeV}} \right)^2
\eeq
and for all the other couplings that only contribute to  $\Delta C_{1u}$ or
to $\Delta C_{1d}$:
\beq
e^2 < .002 \left(\frac{m_{lq}^2}{100 {\rm GeV}} \right)^2 ~~.
\eeq
We have assumed for these constraints that $\lambda_{RS_o} \neq
\lambda_{LS_o}$,
$\lambda_{RS_{1/2}} \neq \lambda_{LS_{1/2}}$,
$\lambda_{RV_o} \neq \lambda_{LV_o}$, and
$\lambda_{LV_{1/2}} \neq \lambda_{RV_{1/2}}$,
because the parity violating coupling is $\lambda_L - \lambda_R$.
This is a reasonable assumption, as noted in \cite{Lang}, because
there are strong bounds on the product of these couplings
$\lambda_L\lambda_R$ from the ratio $(\pi \rightarrow e \nu)/
(\pi \rightarrow \mu \nu)$ (see section 4.1).

\subsection{Neutrino Oscillation Experiments}

Four of the leptoquark couplings induce four-fermion interactions
that couple a neutrino, a charged lepton and an up and down type quark
 (see tables 3 and 4). The couplings $\lambda_{LS_o}, \lambda_{LS_1},
\lambda_{LV_o}$ and $\lambda_{LV_1}$
can therefore be constrained by neutrino oscillation experiments that
look for charged leptons of a different flavour from the original
neutrino beam.

If the neutrino masses are small, and one neglects
the mixing of $\nu_{l_a}$ and $\nu_{l_b}$ with the third neutrino, then the
probability that a neutrino $\nu_{l_a}$ with energy $E$ will oscillate into
a  $\nu_{l_b}$ over a distance $x$ is
\beq
|< \nu_{l_a}|\nu_{l_b}>|^2 = \sin^2 2\theta \sin^2\left(\frac{\Delta m^2 x}
{4E} \right)
\eeq
where $\theta$ is the angle mixing  $\nu_{l_a}$ and  $\nu_{l_b}$, and
$\Delta m^2 = |m_{\nu_a}^2 - m_{\nu_b}^2|$. If
 $\sin^2(\Delta m^2 x/4E)$ oscillates rapidly enough, (large
$\Delta m^2$), it can be averaged to 1/2. The ``large $\Delta m^2$'' bounds
 on the mixing angle $\sin^2 2  \theta$ are therefore bounds on the probability
for a $\nu_{l_a}$ to turn into a $\nu_{l_b}$, or equivalently on the
ratio of $l_a$ to $l_b$ produced by the neutrino beam. This
can be used to constrain the leptoquark induced rate for
$\bar{\nu}_{l_a} + N \rightarrow \bar{l_b} + X ~(\bar{\nu}_{l_a} + u
 \rightarrow \bar{l_b} + [d,s,b])$ and $\nu_{l_a} + N \rightarrow l_b + X
{}~(\nu_{l_a} + u
 \rightarrow l_b + [u,c])$:
\beq
\frac{e_L^2}{m_{lq}^2} < 2 G_F | \sin 2\theta|
\eeq
where $\sin^2 2 \theta$ is from the ``large mass'' upper bound from
accelerator experiments \cite{PDB}.

The scalar leptoquarks that interact with neutrinos have fermion number 2,
so couple  an incoming $\nu$ to an incoming $d$, or an incoming $\bar{\nu}$
to an outgoing $d$. The vectors, on the other hand, have fermion number 0,
and couple an incoming $\nu$ to an outgoing $u$ or an incoming $\bar{\nu}$
to an incoming $u$. This is  important because we assume that the outgoing
quark can be any flavour other than top (we neglect the phase
space suppression due to the mass of the $b$), so to get the flavour indices
right on the couplings constants, we need to know if the neutrino couples
to the leptoquark and the outgoing quark, or the leptoquark and the incoming
quark.

The bounds are listed in the tables.

\subsection{Asymmetries at colliders}

\paragraph{}
The constraints on leptoquarks from the forward-backward asymmetry
($A_{FB}$) at existing $e \bar{e}$ colliders (excluding LEP1)
 were considered by Hewett
and Rizzo \cite{H+R}. (Leptoquarks are not constrained by $A_{FB}$ at
LEP1 because it runs on the $Z$ mass.) Possible
constraints from future colliders were considered
in \cite{DE...}. Hewett and Rizzo required that
$A_{FB}$ and the total cross-section  for $e \bar{e} \rightarrow
q \bar{q}$ not differ from the Standard Model values by more than 5 or
10\% at $\sqrt{s} = 40$ GeV. They claim that a greater contribution
 would disrupt the agreement between Standard
Model predictions and the measured rates for $b$ and $c$ quark
production at PEP and PETRA.

Their bounds apply to the couplings  $(eu)(eu)$ and
 $(ed)(ed)$ ( (11)(11) in the tables), for scalar leptoquarks.
A similar constraint could be calculated for vector leptoquarks,
but since the measurement of $\sin^2 \theta_W$ in
parity violation experiments gives more stringent limits, this
is unneccessary. We list Hewett and Rizzo's bounds in the tables
as
\beq
\lambda^2 < .02 \left( \frac{m_{lq}^2}{100 {\rm GeV}} \right)^2 ~~.
\eeq
Note that this analytic bound is just an approximation to
 the limit graphed in their paper.

In principle, some leptoquarks could contribute strongly
to $A_{LR}$ at tree level, because they only couple to
electrons of one chirality. However, measuring $A_{LR}$
requires polarized electron beams, so we can not get
bounds from this at present (SLC has polarized beams, but, like
LEP1, is running on the Z mass).

\subsection{Muon beams}

\paragraph{}

As was noted by Heusch \cite{Heu}, one should in principle be
able to get interesting bounds on leptoquarks from muon beam experiments.
Leptoquarks contributing to $\mu N \rightarrow \mu X$ can only be constrained
by measurements of asymmetries in this process, because the rate is
dominated by photon exchange. However, if one could effectively
search for $\mu N \rightarrow \ell X$ \cite{H2}, where $\ell$ is an
electron or a tau, one could constrain couplings with the family
structure $\lambda^{\mu q} \lambda^{\ell q'}$, where $q$ and $q'$
can be any type of quark (other than top) if one allows the muon
to scatter off sea quarks in the nucleon.

We can get weak constraints on leptoquarks mediating the
reaction $ \mu q \rightarrow \mu q'$ from the measurement \cite{SPS}
of the asymmetry
\beq
B = \frac{d \sigma(\mu_R^-) - d \sigma( \mu_L^+)}
{d \sigma(\mu_R^-) + d \sigma( \mu_L^+)}
\eeq
where $d \sigma(\mu_R^-)$ is the cross-section for $\mu N \rightarrow
\mu X$ for an incident right-handed muon. If one assumes that only valence
quarks contribute to this process, one can show that
the interference between $Z$ and $\gamma$ exchange gives \cite{SPS,BP}
\beq
B =- \frac{Q^2}{2 \pi \alpha} \frac{G_F}{\sqrt{2}} (c_A^{\mu} - p c_V^{\nu})
(c_A^{d} - 2 c_A^{u}) \frac{6}{5} g(y) \label{asym}
\eeq
where $Q^2$ is the energy transfer, $p$ is the polarization (= .81 in this
experiment), $g(y)$ is a function of the ratio of outgoing to incident muon
energies ($y \equiv 1 - E_f/E_i$), and $c_V^f, c_A^f$ are the usual
vector and axial vector couplings, normalized such that $c_A^{\mu} = -1/2$.
This formula can easily be modified to descibe
the interference between leptoquark
and photon exchange by
substituting $e_L^2/m_{lq}^2 \leftrightarrow 4 \sqrt{2}
G_F$, and setting the various $c$'s to $\pm 1$ or 0 as required. We then
require the leptoquark-photon intererence contribution to  the asymmetry
(\ref{asym}) to be less than the error in $B$. This
gives the bounds listed in the tables.

\section{Discussion}

\paragraph{}
 We have attempted to make this catalogue of constraints as
comprehensive as possible. However, it is inevitable that
we will have omitted some (we have, for instance, neglected
bounds from $CP$ violation).
One could
also do the calculations more carefully than we have done, and
probably improve many of the constraints, as we have tried to err
on the side of caution.

 In the tables, we list the constraints on gauge boson vector leptoquarks.
In some cases, the bounds on non-gauge vectors are much stronger,
and we have tried to estimate these in the text.

  The data in tables 5 to 14 is difficult to interpret, so we
have  collected the best bound on each product
of couplings, and the experiment it came from, in table 15.  The
couplings $\lambda_{LS_o},...\lambda_{LV_1}$ from equations (1) and (2) are
really $3 \times 3$ matrices in generation space. We therefore
 list the matrices horizontally and the indices of the two
matrix elements whose product is constrained vertically. So the top
right-hand entry of  table 15 is the best bound on $\lambda_{LS_o}^{11}
\lambda_{LS_o}^{11}$. As before, the lepton index comes first, and the numbers
in the table are multiplied by $(m_{lq}/100{\rm ~GeV})^2$. We have
assumed, for this table, that all neutrinos are light ($m_{\nu} \ll
m_e$). ALthough we calculated these limits for leptoquarks,
many of the bounds would apply to other massive ``beyond the Standard
Model'' bosons. In particular, the rare meson decay bounds apply
to any particle that induces `$V \pm A$ lepton current' $\times$ `
$V \pm A$ quark current' four-fermion operators.

Ideally we
would like to have individual bounds on the squares of each element
of every  matrix. Instead, we find that the squares of each matrix element
are in general unconstrained. (By this we mean that for a
leptoquark of mass $m_{lq} \sim 100$ GeV, $\lambda^2$ could be of
order 1.) This is hardly surprising, since leptoquark-mediated interactions
whose matrix elements involve the combination
$\lambda^{lq}\lambda^{lq}$ do not change lepton generation or
quark flavour; with two exceptions,  they are therefore competing with
electromagnetic amplitudes for the same process.  For instance,
leptoquarks could mediate the decay of $q \bar{q}$ mesons, but since the photon
can also, there are no interesting bounds on leptoquarks from these
decays. The  two exceptions are
 parity violation experiments, which is where the bounds on
$(\lambda^{11})^2$ come from, and if one of the leptons is
a neutrino. $\lambda_{LS_o}, \lambda_{LS_1}, \lambda_{LV_o}$ and
$\lambda_{LV_1}$ are the
only couplings to involve a neutrino, a charged
lepton and an up and down type quark, and are sometimes better
constrained in consequence.

Most of out constraints come from processes that are suppressed
or forbidden in the Standard Model by lepton family number
conservation, or the absence of flavour-changing neutral currents
among the quarks. Such limits tend to involve the product, rather
than the square, of couplings: $\lambda^{ij}\lambda^{mn} < $
something, with
$(ij) \neq (mn)$. An upper bound on the product of two couplings is
not terribly useful, because either coupling could be very large or
very small.
   The other problem associated with bounds from flavour changing
processes is that usually $i \neq j$ and/or $m \neq n$ in bounds of
the form $\lambda^{ij} \lambda^{mn} < $ something.

If the leptoquark couplings
have a generation structure similiar to the Standard Model, we might
expect couplings $\lambda^{ij}$ with $i \neq j$ to be one or two orders of
magnitude smaller than the $i = j$ value, by analogy with the CKM
matrix. It is unfortunate that the couplings that we might expect to
be largest are the ones we have the fewest constraints on.

 We have few bounds on leptoquark couplings to the top quark,
because these can only be obtained  from loop diagrams
(meson-anti-meson mixing and  $\mu$ and $\tau$ decay to three leptons
or a lepton and a photon), and the quark mass was neglected in the
triangle loop calculations. This could be corrected, but would still
not lead to simple numerical bounds, because they would depend on the
unknown leptoquark and top masses. The interactions
which are unconstrained because they involve a top
quark are indicated by stars in table 15.

It is worth noting that some of the interactions couple a charged lepton
to an up-type quark, some to a down, and some to both. The limits
on the
coupling constants of interactions involving down-type quarks
tend to be better than those for  ups, because $K$s are better
constrained than $D$s, and $B$s exist.

  HERA expects to be able to see leptoquarks with $m_{lq} \lappeq$
300 GeV and $\lambda^2 \gappeq 10^{-4}$ as peaks in the
$e^{\pm} p \rightarrow e^{\pm} X$ cross-section as a function of
$x$ \cite{BRW}, so if  the upper
bound on $\lambda_{\alpha}^{11} \lambda_{\alpha}^{mn}$ from a
rare decay is greater
than $10^{-5} (m_{lq}/100{\rm ~GeV})^2$, then the leptoquark
`$\alpha$' can be produced and detected via these
couplings at HERA.
As is clear from table 15, this means that it is consistent with
the  bounds computed here for HERA to see any
of the leptoquarks.  Four of the constraints listed here
are strong enough to each rule out a particular production
and decay channel, but since the leptoquark could easily decay
to a different quark-lepton final state, this is not very significant.
For instance, the bounds from muon conversion on
titanium ($e^2 < 7 \times 10^{-7} (m_{lq}/100{\rm ~GeV})^2$) suggest
that no leptoquark produced in the collision  of an electron (or
$\bar{e}$) and a first generation quark will be detected decaying to
a muon and a first generation quark (a second or third
generation quark would, however, be
quite possible). Similiarly, the absence of kaon
decays to $\bar{\mu} \mu, \pi \mu \bar{e}$  or $e \bar{\mu}$ implies
that HERA will not see the leptoquarks $V^{\mu}_o, V^{\mu}_{1/2},
\vec{V}^{\mu}_1, \tilde{S}_o, \tilde{S}_{1/2}, {\rm ~and~} \vec{S}_1$
decaying to a strange quark and an
electron or muon (but again, some other flavour of quark
would not be ruled out).  If the leptoquark Yukawa couplings were generation
{\it independant}, these constraints would apply to all initial and
final quark-lepton combinations, and HERA would not see leptoquarks.
However, if one makes the far more reasonable assumption that the
couplings are generation dependent (very true of the Standard Model
fermion-fermion-boson couplings), then our bounds from rare
processes  do not
seriously infringe on HERA's chances of seeing leptoquarks. The
 experiments listed above  do constrain
$e^2/m_{lq}^2 < 10^{-9}$ GeV$^{-2}$ ($e^2 <
10^{-5}$ in the tables), but they do not apply to
the couplings for $e^{\pm} p \rightarrow e^{\pm} X$ (where $X$ consists
of $u$ and $d$ quarks).
So leptoquarks can easily  have masses and
couplings that are accessible to HERA and consistent with the upper
bounds on rare processes.

In figure 5 we have plotted, following \cite{BRW} and using quark
distributions from \cite{MT}, the differential cross-section
for $e^- p \rightarrow e^- X$ in the presence of the
scalar leptoquark $S_o$ with a mass of 200 GeV, and couplings
$\lambda_{LS_o}^{11} = .008, \lambda_{LS_o}^{ij} = 0$ for $i,j \neq 1$, and
$\lambda_{RS_o} = 0$.
This choice of mass and coupling is consistent with  the upper bound
$(\lambda_{S_o}^{11})^2 < 2 \times 10^{-3} (m_{lq}/100~{\rm GeV})^2$ from
quark-lepton universality. One could make
similiar plots for other leptoquarks, and second or third
generation final state quarks and leptons. It is clear that
 if ZEUS and
H1 do not see leptoquarks, they will impose
strong constraints on all couplings of the form $\lambda^{11}\lambda^{mn}$.

\section{Conclusion}

\paragraph{}
We have calculated generation dependent upper bounds on the
coupling constants of all renormalizable, $B$ and $L$ conserving
interactions consistent with the $SU(3) \times SU(2) \times U(1)$ symmetry
of the Standard Model, involving a lepton, a quark, and a scalar
or vector leptoquark. The constraints come from rare meson
decays, meson-anti-meson mixing, lepton decays, and a miscellany
of electroweak tests. We list the bounds on the square of each
coupling constant from each experiment in tables 5 through 14;
since the constraints depend on the leptoquark mass, and only apply
for certain generation indices, we quote numerical bounds based on
$m_{lq} = 100$ GeV, and put the lepton and quark generation indices
in parentheses ( so that the top right hand corner of table 5:
$(11)(n1) < 10^{-3}$ means that $\lambda_{LS_o}^{ed} \lambda_{LS_o}^{\nu^n u} <
10^{-3} (m_{lq}/100 {\rm GeV})^2$). In table 15, we list the
best bound on each coupling constant for every combination of generation
indices.

\section*{acknowledgements}

\paragraph{}
We are very grateful to Doug Gingrich and  Nathan Isgur for their
extensive help, and  would like to thank Andrzej Czarnecki, Xiao-Gang He,
JoAnne Hewett, Ian Hinchliffe, Drew Peterson, Tom Rizzo
and Martin White for
useful conversations. This research
was partially supported by the Natural Sciences and Engineering Research
Council of Canada.


\begin{thebibliography}{222222}
\bibitem{LQHERA} for a review of leptoquarks at HERA, see:
 R.J. Cashmore et al., {\it Phys. Rep.} {\bf 122} (1985) 275 \\
 ``{\it Physics at HERA}'', Proceedings of the Workshop, Hamburg, 1991, W.
Buchmuller, G. Ingelman, editors.
\bibitem{BRW} W. Buchmuller, R. Ruckl, and D. Wyler, {\it  Phys.
Lett.} {\bf B191}  (1987) 442.
\bibitem{ML2} M. Leurer, {\it Phys. Rev.}  {\bf D46}  (1992) 3757.
\bibitem{virlq} J.A.Grifolis, S. Peris, {\it Phys. Lett}
 {\bf B201} (1988) 287.\\
M.A. Doncheski, J.L. Hewett, {\it Zeit. Phys} {\bf C56} (1992) 209.
\bibitem{Heu} C.A. Heusch, lecture at the ``Rencontre de physique de
la vall\'{e}e d'Aoste'', February 1989.
\bibitem{D+E} S. Dimopoulos, J. Ellis, {\it Nucl. Phys} {\bf B182} (1981) 505.
\bibitem{S1} O. Shenkar, {\it Nucl. Phys.} {\bf B206} (1982) 253.
\bibitem{S2} O. Shenkar, {\it Nucl. Phys.} {\bf B204} (1982) 375.
\bibitem{MSW} R. Mohapatra, G. Segr\'e, L. Wolfenstein, {\it Phys. Lett}
{\bf B145} (1984) 433.
\bibitem{BKZ} I. Bigi, G. K\"{o}pp, P.M. Zerwas, {\it Phys.
Lett.} {\bf  166B} (1986) 238.
 \bibitem{BW} W. Buchmuller and D. Wyler, {\it  Phys. Lett.} {\bf
B177}  (1986) 377.
\bibitem{Leff} W. Buchmuller and D. Wyler, {\it  Nucl. Phys. } {\bf
B268}  (1986) 621.
\bibitem{CEEGN} B.A. Campbell, J. Ellis, K. Enqvist, M.K. Gaillard,
and D.V. Nanopoulos, {\it Int. J. Mod. Phys. } {\bf A2} (1987) 831.
\bibitem{E6} J. Hewett, T. Rizzo, {\it Phys. Rep.} {\bf 183} (1989)  193.
\bibitem{D+H} A.J.Davies, X. He, {\it Phys. Rev.} {\bf D43} (1991) 225.
\bibitem{HMP} X. He, B.H.J. McKellar, S. Pakvasa, {\it Phys. Lett.}
{\bf B283} (1992) 348.
\bibitem{G} C.Q. Geng, {\it Zeit. Phys.} {\bf C48} (1990) 279.
\bibitem{BF} S.M. Barr, E.M. Freire, {\it Phys. Rev.} {\bf D41} (1990) 2129.
\bibitem{Ross} G.G. Ross, {\em Grand Unified Theories}, Benjamin/Cummings,
1984.
\bibitem{P+S} J.C. Pati, A. Salam, {\it Phys. Rev.} {\bf D8} (1973) 1240;
{\it Phys.Rev. Lett.} {\bf 31} (1973) 661;
{\it Phys. Rev.} {\bf D10} (1974) 275.
\bibitem{Lan} P. Langacker, {\it Phys. Rep.} 72 (1981) 187.
\bibitem{G+G} the first SU(5) model was proposed by H. Georgi, S.L. Glashow,
{\it Phys. Rev. Lett.} {\bf 32}  (1974) 438.
\bibitem{M+Y} H. Murayama, T. Yanagida, TU 370 (1991).
\bibitem{TC} for a review, see G.G. Ross, {\it Grand Unified Theories},
Benjamin-Cummings, 1985 \\ E. Farhi, L. Susskind, {\it Phys. Rep.} {\bf 74 }
(1981) 277 \\
E. Eichenten et al., {\it Rev. Mod. Phys.} {\bf 56} (1984) 579.
\bibitem{subrev} for a review, see: I. Bars, {\it Proc. of the 1984
Summer Study on the SSC},
editors R.Donaldson, J. Morphin (APF New York, 1985) 38. \\ W. Buchm\"uller,
{\it Acta Phys. Austr. Suppl.} {\bf  XXVII} (1985) 517. \\
B. Schrempp, {\it Proc. of the XXIII
 Int. Conf. on High Energy Physics}, Berkeley, \\ R. Peccei, in
{\it Proceedings of the 2nd Lake Louise Winter Institute},
 J.M. Cameron et al., eds.,
World Scientific, Singapore (1987).
\bibitem{AF} L.F. Abbott, E. Farhi, {\it Phys. Lett.} {\bf 101B}
(1981) 69; {\it Nucl. Phys.}
{\bf B189} (1981) 547.
\bibitem{SUSYC} W. Bardeen, V. Visnji\'c, {\it Nucl. Phys.} {\bf  B194}
 (1982) 151. \\ W. Buchmuller, R. Peccei, T. Yanagida, {\it Phys. Lett.} {\bf
 124B} (1983) 67. \\R. Barbieri, A. Masiero, G. Veneziano,
{\it Phys. Lett.} {\bf 128B} (1983) 179.
\bibitem{SS} B. Schrempp, F. Schrempp, {\it Nucl. Phys.} {\bf B231} (1984) 109.
\bibitem{other} J.C. Pati, {\it Phys. Lett. } {\bf B228} (1989) 228. \\
 see also references in \cite{subrev} for other composite models.
\bibitem{'tH} G. 't Hooft, in {\it Recent Developements in Gauge Theories}
ed. G 't Hooft et al., Plenum, N.Y. (1980).
\bibitem{Frish} Y. Frishman et al., {\it Nucl.  Phys. } {\bf B177}
(1981) 157.
\bibitem{C+G} S. Coleman, B. Grossman, {\it Nucl. Phys.} {\bf B203}
(1982) 205.
\bibitem{DRS1} S. Dimopoulos, S. Raby, L. Susskind, {\it Nucl. Phys. }
 {\bf B169} (1980) 373.
\bibitem{DRS2}S. Dimopoulos, S. Raby, L. Susskind, {\it Nucl. Phys. }
 {\bf B173} (1980) 208.
 \bibitem{Peskin} M. Peskin, in {\it Proceedings of the 1985 Lepton-Photon
Symposium}.
\bibitem{CGWW} K.M. Case, S.G. Gasierowicz, {\it Phys. Rev.} {\bf 125}
(1962) 1055.
S. Weinberg, E. Witten, {\it Phys. Lett. } {\bf 96B} (1980) 159.
\bibitem{H+R} J.L.  Hewett, T.G.  Rizzo, {\it Phys. Rev.} {\bf D36}
(1987) 3367.
\bibitem{PDB} $\underline{\rm Particle ~Data ~Book}$, {\it Phys.
Rev.} {\bf D45}, \#11 (1992).
\bibitem{subLQ} W. Buchmuller,{\it  Phys. Lett.} {\bf 145B}
 (1984) 151. \\ B. Schrempp, F. Schrempp, {\it Phys. Lett.} {\bf 153B}
 (1985) 101.
\bibitem{AMY} AMY Collaboration, G.N. Kim et al., {\it Phys. Lett.}
{\bf B240} (1990) 243.
\bibitem{L3} ALEPH Collaboration, CERN PPE 91/149. \\ L3
Collaboration, {\it Phys. Lett.}  {\bf B261} (1992) 169,\\ OPAL
Collaboration,    {\it Phys. Lett.}  {\bf B263} (1992) 123.
\bibitem{Gen} M. A. Doncheski et al., {\it Phys. Rev.} {\bf D40}
(1989) 2301.
\bibitem{DDPF} T.D Papadopoulos,, for the Delphi Collaboration, in
{\it Proceedings of the Fermilab Meeting, DPF '92}, World Scientific.
\bibitem{UA2} UA2 Collaboration, {\it Phys. Lett.} {\bf B274} (1992)
507.
\bibitem{Carl}  CDF Collaboration, Abe et al., Fermilab-Pub-93/070-E,
submitted to {\it Phys Rev} {\bf D}.
\bibitem{D0} D0 Collaboration, talk presented at
 the 1993 Lepton Photon Symposium.
\bibitem{J+T} J. Hewett, S. Pakvasa, Pomeral, T. Rizzo, work in progress.
\bibitem{ZEUS} ZEUS Collaboration, {\it Phys. Lett.} {\bf B306}
(1993) 173. \\ H1 Collaboration, {\it Nucl. Phys. } {\bf B396} (1993) 3. \\
K. McLean for the ZEUS Collaboration, at the ``Conference
of the European Physical Society'', Marseille, 1993.
\bibitem{Rexpt} D. I. Britton et al., {\it Phys. Rev. Lett.} {\bf 68} (1992)
3000. \\G. Czapek et al., {\it Phys. Rev. Lett.} {\bf 70} (1993)
17. \\ We get the numbers quoted in the text by averaging
 the results and the errors.
\bibitem{Rth} S. Berman, {\it Phys. Rev. Lett.} {\bf 1} (1958) 468.\\
  T. Kinoshita, {\it Phys. Rev. Lett.} {\bf 2} (1957) 477.  \\
W.J. Marciano, A. Sirlin, {\it Phys. Rev. Lett.} {\bf 36} (1976)
1425.  (1.233 $\pm .004 \times 10^{-4}$). \\
T. Goldman and W. Wilson, {\it Phys. Rev.} {\bf D15} (1977) 709,
 (1.239 $ \pm .001 \times 10^{-4}$). \\
W. Marciano, see \cite{Rexpt}, (1.2345 $\pm .0010 \times 10^{-4})$.
\bibitem{pi0} A. Deshpande et al., {\it Phys. Rev. Lett.} {\bf 71} (1993)
27. ($6.9 \pm 2.3 \pm .6 \times 10^{-8}$) \\
K.S. McFarland et al., {\it Phys. Rev. Lett.} {\bf 71} (1993)
31. ($7.6 +3.9 - 2.8  \pm .5 \times 10^{-8}$)
 \bibitem{Mart}G.Martinelli in {\it Proceeedings of the XXVth Rencontre
de  Moriond: Electroweak Interactions and Unified Theories},
Editions Fronti\`{e}res, 1991.
\bibitem{ML} M. Leurer, Weizmann Institute preprint, WIS-93/26/March-PH.
\bibitem{MW} for a review, see M. Wise, in {\it Proceedings of the
6th Lake Louise Winter Institute} (1991), eds. B.A.Campbell et al., World
Scientific, Singapore.
\bibitem{HEPdal} ALEPH Collaboration, contributed paper at
the XXVI International Conference on High Energy Physics, Dallas, Texas (1992).
\bibitem{Vub} R. Fulton et al., CLEO Collaboration,
 {\it Phys. Rev. Lett.} {\bf 64} (1990) 16. \\
H. Albrecht et al., ARGUS Collaboration,
 {\it Phys.  Lett.} {\bf B255} (1991) 297.
 \bibitem{CLEOt} CLEO Collaboration, {\it Phys. Rev.} {\bf  D35} (1987) 3533.
\bibitem{UA1} UA1 Collaboration, {\it Phys. Lett.} {\bf B262} (1991) 163.
\bibitem{IW} N. Isgur, M. Wise {\it Phys. Rev.} {\bf D42} (1990) 2388.
\bibitem{ISGW} N. Isgur, D. Scora, B. Grinstein, M. Wise, {\it Phys.
Rev.} {\bf D39} (1991) 799.
\bibitem{fD} J.C. Anjos et al., {\it Phys. Rev. Lett.} {\bf 67} (1991) 1507.
\bibitem{eq} Mark III Collaboration, J. Adler et al., {\it Phys. Rev. Lett.}
{\bf 62} (1989) 1821.
\bibitem{BKg} CLEO Collaboration, CLNS-93-1212.
\bibitem{D} H. Dreiner, {\it Mod. Phys. Lett} {\bf A3} (1988) 867.
\bibitem{Drew} K.A. Peterson, {\it Phys. Lett.} {\bf  B 282 } (1992) 207.\\
K.A. Peterson, Ph.D. thesis, University of Alberta, 1991.
\bibitem{Bry} S. Ahmad et al., {\it Phys. Rev. Lett.} {\bf 59}
(1987) 970.\\
 S. Ahmad et al., {\it Phys. Rev.} {\bf D38} (1988) 2102.
\bibitem{S} O. Shenkar, {\it Phys. Rev. } {\bf D 20} (1979) 1608.
\bibitem{C+H} R.N. Cahn, H. Harari, {\it Nucl. Phys. } {\bf B176} (1980), 135.
\bibitem{C+L} T.Cheng, L. Li, $\underline{\rm Gauge ~Theory ~of
{}~Elementary ~Particle ~Physics}$, Oxford University Press, (1989)
chapter 12.
 \bibitem{G+L} M.K. Gaillard, B.W. Lee, {\it Phys. Rev. } {\bf D10}
(1974) 897.
 \bibitem{RND} B.W. Lee, R.E.  Shrock, {\it Phys. Rev. } {\bf D16}
(1977) 1444.
\bibitem{CLEO} CLEO collaboration, {\it Phys. Rev. Lett.} {\bf 70} (1993) 138.
\bibitem{ARGUS} ARGUS Collaboration, {\it Zeit.
Phys.} {\bf C55} (1992) 179.
\bibitem{Mor} D.A. Morris, {\it Phys. Rev.} {\bf D 37} (1988) 2012.
\bibitem{Lev} J.P. Leveille, {\it Nucl. Phys.} {\bf B137} (1978) 63.
\bibitem{DKST} A. Djouadi, T. Kohler, M. Spira, J. Tutas, {\it Zeit.
Phys.} {\bf C46} (1990) 679.
\bibitem{HePC} X. He, private communication.
\bibitem{MS} W.J. Marciano, A.I. Sanda, {\it Phys. Lett.} {\bf 67B} (1977) 303.
\bibitem{Spira} M. Spira, diploma thesis, PWTH Aachen, 1989.
\bibitem{QCD} B. Grinstein, R. Springer, M.B. Wise, {\it Phys. Lett.}
{\bf  B 202} (1988) 138; {\it Nucl. Phys} {\bf B339} (1990) 269.
\\ R. Grigjanis, P.J. O'Donnell, M. Sutherland, H. Navelet,
{\it Phys. Lett.} {\bf  B 223 } (1989) 239.
\bibitem{bsg} B.A. Campbell, P.J. O'Donnell, {\it Phys. Rev.} {\bf  D 25}
 (1982) 1989.\\
S.P. Chia, {\it Phys. Lett.} {\bf 240B} (1990) 465.\\
T. Inami, C.S. Lim, {\it Prog. Theor. Phys.} {\bf 65} (1981) 297.
\bibitem{Her} P. Herczag, in {\it Fundamental Symmetries in
 Nuclei and Particles}, World Scientific, 1989.
\bibitem{Lang} P. Langacker, {\it Phys. Lett.} {\bf 256B} (1991) 277.
\bibitem{H2} C.A. Heusch, in {\bf Progress in Electroweak Interactions},
J. Tran Thanh Van, ed., (1986).
\bibitem{DE...} H. Dreiner et al., {\it Mod. Phys. Lett} {\bf A3} (1988) 443.
\bibitem{SPS} A. Argento et al., {\it Phys. Lett. } {\bf B120} (1983) 245.
\bibitem{BP} S.M. Berman, J.R. Primack, {\it Phys. Rev.} {\bf D 9} (1974) 2171.
\bibitem{MT} J.G. Morfin, W.K.  Tung, {\it Zeit.
Phys.} {\bf C52}  (1991) 13.
\end{thebibliography}
\end{document}